\documentclass[preprint, 12pt,5p,times]{elsarticle}
\usepackage{amsmath,amssymb,amsfonts}
\usepackage{algorithmic}
\usepackage{graphicx}
\usepackage{multirow}
\usepackage{float}
\usepackage{amsmath}
\usepackage{mathtools}
\usepackage{amsmath}
\usepackage{nccmath}
\usepackage{amssymb}
\usepackage{hyperref}
\usepackage{sidecap}
\usepackage{wrapfig}
\usepackage{subfig}
\usepackage{tabularx}

\usepackage[none]{hyphenat}
\journal{Neural Networks}

\begin{document}
	
\begin{frontmatter}
	
	
	
	\title{Associated Random Neural Networks for Collective Classification of Nodes in Botnet Attacks
	\tnotetext[t1]{This research was supported by the IoTAC Research and Innovation Action,
		funded by the European Commission (EC) under H2020 Call SU-ICT-02-2020 ``Building blocks for resilience in evolving ICT systems'', under Grant Agreement No. 952684.}}
	

	\author[inst1,inst2,inst3]{Erol Gelenbe\corref{cor1}}
	\ead{seg@iitis.pl}
	\cortext[cor1]{Corresponding author}
		\author[inst1]{Mert Nak\i p}
	\ead{mnakip@iitis.pl}
	
	\address[inst1]{Institute of Theoretical and Applied Informatics,
		Polish Academy of Sciences (PAN), 
		Bałtycka 5,
		Gliwice,
		44--100, 
		Poland
	} 
	\address[inst2]{Lab. I3S, Universit\'{e} C\^{o}te d'Azur, 
		Cedex 2,
		Nice,
		06103, 
		France
	}
	\address[inst3]{Ya\c{s}ar University, 
	Bornova,
	Izmir,
	35100, 
	Turkey
	}
	
	\begin{abstract}
	Botnet attacks are a major threat to networked systems because of their ability to 
	turn the network nodes that they compromise into additional attackers, leading to the spread of high volume attacks over long periods. The detection of such Botnets is 
	complicated by the fact that multiple network IP addresses will be simultaneously compromised,
	so that Collective Classification of compromised nodes, in addition to the already available traditional methods that focus on individual nodes, can be useful. Thus this work introduces a collective Botnet attack classification technique that operates on traffic from
	a $n$-node IP network, with a novel 
	Associated Random Neural Network (ARNN) that identifies the nodes
	which are compromised. The ARNN is a recurrent architecture that  incorporates two mutually associated, interconnected and architecturally identical $n$-neuron random neural networks,
	that act simultneously as mutual critics to reach the decision regarding which of $n$ nodes have been compromised. A novel gradient learning descent algorithm is presented  
	for the ARNN, and is shown to operate effectively both with conventional off-line training from prior data, and with on-line incremental training
	without prior off-line learning.  Real data from a $107$ node packet network is used
	with over $700,000$ packets to evaluate the  ARNN, showing that it provides accurate predictions. Comparisons with other well-known state of the art methods using the same learning and testing datasets, show that the ARNN offers significantly better performance.
	\end{abstract}
	
	
	%	\item Research highlight 1
	%	\item Research highlight 2
	
	\begin{keyword}
Collective Classification \sep Botnet Attack Detection \sep Associated Random Neural Networks \sep	The Internet \sep Nodes Compromised by Botnets \sep Random Neural Networks \sep ARNN Learning  
	\end{keyword}
	
\end{frontmatter}

\section{Introduction} \label{section:introduction}

Many classification problems, such as identifying a given individual's face in a large
dataset of face images of people \cite{Davis}, associate a binary label to data items \cite{Binary}. This is also the usual case for network attack detection from traffic data \cite{Botnet1} that attemps to determine
whether a given network node has been compromised by an attack \cite{Filus21}. Such problems are often solved with Machine Learning (ML) algorithms that learn off-line from one or more datasets that contain the ground-truth data. The trained ML algorithm can then be tested on datasets that have not been used for learning, and then used online with previously unseen or new data. Typically, the online usage of such attack detection algorithms is carried out ``one node at a time'', i.e. as an individual classification problem for a specific node that may be concerned by possible attacks \cite{Alshamkhany,Access22}.

When we need to classify each individual node in a set $V=\{v_1,~...~,v_n\}$ of interconnected nodes in a network as being ``compromised'' or uncompromised (i.e.,  ``safe'') we obviously face with a Binary {\em Individual} Classification Problem for each of the $n$ nodes. However, when the attacking entity is a Botnet which induces a compromised node to attack several other nodes with which it is able to directly communicate, then we are faced with a {\em Collective Binary Classification Problem}
where the classification of the distinct nodes is correlated, even though we cannot be sure that
a compromised node has sufficient bandwidth or processing capacity to actually compromise other nodes. 

Indeed  let $A=[A_{ij}]_{n\times n}$ be the (deterministic)  adjacency matrix
where $A_{ij}=1$ indicates that node $v_i$ has opened a connection to node $v_j$ and therefore can send packet traffic to it, while $A_{ij}=0$
indicates that node $v_i$ is unable to send packets to node $v_j$. Then  during a Botnet attack, nodes that can receive traffic from compromised nodes are themselves likely to become compromised, and to become in turn attackers against other nodes, so that one needs to classify nodes by taking account both the local atack traffic at each node, and their patterns of communication between nodes. 

Collective (also known as ``relational'') classification problems have been widely studied \cite{Collective1,Collective3} using a variety of  techniques
linked to ML.  As indicated in the literature \cite{Collective2},
collective classification may use a collection of local conditional classifiers which 
classify an individual's label conditionally on the label value of others, and then fuses
the overall sets of outputs, or may try to
solves the problem either as a global optimization or
a global relaxation problem \cite{Collective4,Collective5}, with the global approach being
often computationally more costly.

Botnet attack detection has been discussed in numerous papers, mainly using 
single node attack detection techniques \cite{CollectiveSurvey1,CollectiveSurvey2,CollectiveSurvey3} which can identify
individually comprimised nodes, except for some studies that analyze relations between nodes to detect the existence or spread of a Botnet \cite{CollectiveDetect1,CollectiveDetect2,CollectiveDetect3}. 

Thus in this paper we address the Collective Classification problem of detecting all the nodes in a given network which have been
compromised by a Botnet.  In particular, we introduce a ML method that combines
supervised learning by a novel Random Neural Network \cite{RNN1} architecture -- which we call the Associated Random Neural Network (ARNN) -- that learns from a sample taken from the traffic flowing among a set of network nodes, to classify them as being either compromised by a Botnet, or as non-compromised.  

The Random Neural Network is a bio-inspired spiking
Neural Network that has a convenient mathematical solution, and  has been applied by numerous authors, including \cite{Video,Aiello,Khaled0,Kaptan1,Khaled1,Khaled2,BuildingEnergy2,VideoScheduling,Intrusion,AdaptiveModulation,HVAC1,HVAC2,BuildingEnergy1,VoiceQuality,VideoQuality,Virus,Rubino1,Rubino2,Rubino3,Yin1,Yin4}, in diverse problems that can be addressed with ML such as video compression, tumor recognition from MRI images, video quality evaluation, smart building climate management, enhanced reality, voice quality evaluation over the internet, wireless channel modulation, climate control in buildings, the detection of network viruses and other cyberattacks.

In the case of Botnet detection, the   ARNN is trained off-line with data that is certified as containing Botnet attacks, and with data that is attack free, and the trained   ARNN is then used online to monitor a network's traffic to collectively classfy  which nodes -- if any -- are compromised by a Botnet.

In the sequel, Section \ref{Survey} surveys previous  research on Botnet attacks. In Section \ref{Method} the proposed ARNN is described; to improve readability its gradient learning algorithm is detailed separately in \ref{Appendix}. 

Section \ref{Experimental} presents the experimental work based on a large MIRAI Botnet dataset involving $107$ network nodes and over $760,000$ packets \cite{KitsuneKaggle} that is used for training and evaluating the proposed method. The evaluation of the  ARNN using this dataset is detailed in Section \ref{Eval}, where we have also compared our results with other well known ML methods. Finally, conclusions and suggestions for further work are presented in Section \ref{Conclusions}.

\section{Recent Work on Botnet Attack Detection}\label{Survey}

In networked systems the cost of not meeting security requirements can be very high \cite{bib:ciscopriv,bib:cisco,HTTP}, hence much effort has been devoted to developing techniques that {\bf detect attacks} against network components such as hosts, servers, routers, switches, IoT devices, mobile devices and various network applications.

Botnet attacks are particularly harmful, since they induce their victims to become
sources of further attacks  against third parties
\cite{Survey,Botnet25,Botnet2}. Recent Botnet reports include the 2016 MIRAI attack \cite{Mirai}, and the MERIS type attacks from 2021 and 2022 that can generate some $46$ million requests per second, lasting more than $60$ minutes, exploiting over $5,000$ source IP addresses as Bots from over $130$ countries
\cite{Meris1,Meris2}, which is a similar rate of requests as all the Wikimedia daily requests 
made in ten seconds. Another MERIS attack generated $17.2$  million requests per second against a commercial  web site, and such attacks have been observed to
target some $50$ web sites per day, with over $100$ Distributed Denial of Service (DDoS) attacks,
of which one third appear to occur in China, and $13\%$ in the USA,
involving a number of Bots sometimes ranging between $30,000$ up to  $250,000$.

Botnet attack detection techniques typically  
examine incoming traffic streams and identify sub-streams  that are benign or ``normal'', and those that may contain attacks \cite{bib:Jeatrakul2009,Botnet12,Botnet8}, and
often classify  attacks into ``types''  \cite{bib:yin2017deep} based on signatures \cite{bib:cortes2019hybrid, bib:li2019designing} that exploit prior knowledge about attack patterns. In addition, false alarms should also be minimized
so that useful network traffic is not eliminated by mistake. However, such methods can also be overwhelmed by attack generators \cite{bib:medbiotguerra2020medbiot} that have been designed to adaptively modify their behaviour.

 Defense techniques for Botnets based on the smart location of counter-attacks by ``white hat'' worm launchers have also been suggested \cite{Botnet3,Botnet7},
while refined deep learning (DL) techniques have been investigated to recognize constantly evolving 
Botnet traffic \cite{Botnet4}, and transfer learning can improve detection accuracy, without concatenating large datasets having different characteristics \cite{Botnet5}. 

Recent work has also created a taxonomy of Botnet communication patterns including encryption and hiding \cite{Botnet6} with some authors examining how Internet Service Providers (ISP) can participate collectively to mitigate their effect \cite{Botnet19}. Other work suggests that traditional Botnet detection techniques in the Internet are not well adapted to emerging applications such as the IoT \cite{Botnet10}, some studies have addressed Botnet apps in specific operating system contexts such as Android \cite{Botnet17} or Botnet detection for specific applications such as Peer to Peer Systems (P2P) \cite{Botnet22}, or Vehicular Networks for which specific detection and protection mechanisms are suggested \cite{Botnet21}. 

Some recent research has  focused on the manner in which Botnet variability can be reflected in intrusion detection software that is designed for a given host \cite{Botnet24}. Universal sets of features that may be applicable to attack detection \cite{Botnet26} have also been
suggested, and detection techniques for specific types of Botnets such as the ones based on the Domain Generation Algorithm \cite{Botnet23} have been proposed.

Most of the previous literature on Botnets, as well as our recent work, has focused on {\em single node detection} with off-line learning. We developed detection techniques for Distributed Denial of Service (DDoS) attacks
using {\em gradient descent learning} with the RNN \cite{Filus21,Spilios}, because Botnets often use DDoS as the means of bringing down their victims. 
The system-level remedial actions that should be taken after an attack is detected \cite{Gelenbe2020} were also analyzed.  To avoid learning all possible types of attack patterns, an {\em auto-associative approach  based on Deep Learning of ''normal‘‘ patterns with a dense multi-layer RNN} \cite{gelenbe2017deepdense} was developed to detect malicious attacks by identifying deviations from normal traffic \cite{Brun, Nakip, Nakip_incremental}.  It was also shown that a single trained auto-associative dense RNN
can provide detection of multiple types of attacks (i.e. not just Botnets) \cite{G-Nets-Attack},
and that learning can be partially conducted on-line, with less need for long and computationally costly off-line training \cite{Access22}.

\subsection{Approach Developed in this Paper}

While it is possible to accurately detect malicious attacks by processing traffic at a given node, it is difficult to certify that the detected attack is indeed a Botnet by observing a single node since Botnets are based on the propagation of attack patterns through multiple nodes. Furthermore, many attack detectors detect anomalies in the incoming traffic rather that pointing to a specific attack \cite{G-Nets-Attack}. Thus the present paper develops a Collective Classification approach to secifically address the Botnet detection problem in the following manner:
\begin{enumerate}
	\item A finite set of $n$ interconnected network IP (Internet Protocol) addresses is considered,
	\item Some of these addresses are equipped with a Local Attack Detector (LAD), so that a local evaluation is available at some of the nodes about whether they are being attacked. Note that the fact that a node is attacked  does not necessarily imply that it has been compromised,
	\item A specific neural network architecture, the Associated RNN (ARNN) with $2n$ neurons, is designed {\bf to deduce  which (if any) of the IP addresses have been {\em compromised}}  by Botnet(s), using the available decisions from the LADs regarding individual nodes. The ARNN is trained,
	using the algorithm detailed in \ref{Appendix}, on a small  subset of data taken from a large open access Botnet dataset \cite{KitsuneKaggle} containing over $760,000$ packets exchanged among  $107$ IP (Internet Protocol) addresses.
	\item Then using the remaining large dataset (not used for training) we determine which of the $107$ IP addresses have been compromised and become Botnet attackers, resulting in a high level of accuracy regarding which IP addresses are compromised.
	\item Two other well established ML methods are also used to identify which of the $107$ nodes have been compromised. The results show  that the  ARNN provides significantly better accuracy concerning both True Positives and True Negatives.
	\end{enumerate}

\section{The  ARNN Decision System}
\label{Method}

\begin{figure}[h!]
	\centering 

	\includegraphics[height=10cm,width=8.5cm]{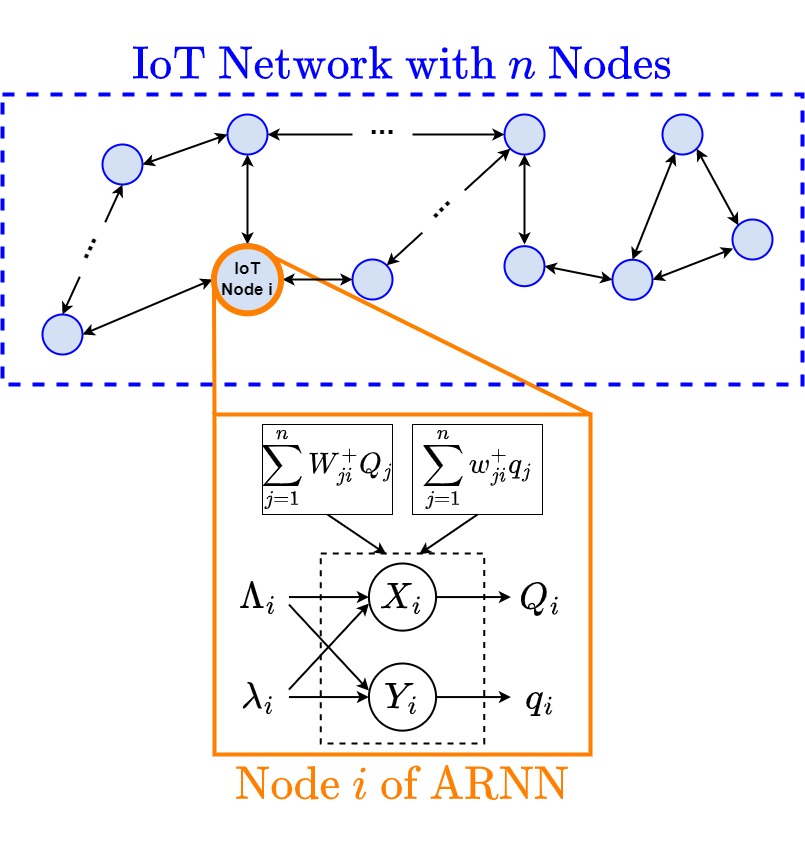}
	\caption{A schematic diagram of  the $2n$-neuron  ARNN that carries out a Collective Classification of the compromised nodes (if any) for a 	$n$-node IoT network denoted $V=\{v_1,~...~,v_n\}$.  The ARNN has two neurons $X_i$ and $Y_i$ that represent opposite views for each network node $v_i$: $X_i$ indicates that $v_i$ is compromised, while $Y_i$ indicates that $v_i$ is not compromised.  The corresponding numerical decision variables are $Q_i,~q_i\in [0,1]$, where $Q_i$ is the probability that $X_i$ is excited and $q_i$ is the probability that $Y_i$ is excited. $X_i$ has an excitatory connection $W^+_{ij}$ to each other neuron $X_j$ and an inhibitory connection $W^-_{ij}$ to all other $Y_j$ neurons, and $Y_i$ has an excitatory connection $w^+_{ij}$ to each other neuron $Y_j$ and an inhibitory connection $w^-_{ij}$ to all other $X_j$ neurons. A neuron does not excite or inhibit its own self. Thus inside the  ARNN, the neurons of type $X$ excite other neurons of type $X$ and inhibit all neurons of type $Y$, and vice-versa for the neurons of type $Y$. The ARNN is ``self-critical'' in the sense that neurons of type $X$ try to supress the neurons of type $Y$, and vice-versa. $\Lambda_i$ represents the output from the LAD (local attack detector) at node $v_i$ stating that $v_i$ has been compromised while $\lambda_i$ represents the LAD output at node $v_i$ stating that it has not been compromised. $\Lambda_i,~\lambda_i$ act as an excitatory and inhibitory external input, respectively,  for $X_i$, while they act as an inhibitory, excitatory input for $Y_i$.}
	\label{fig:SystemDesign}
\end{figure}

The decision system presented in this paper, the ``self-critical'' ARNN with $2n$ neurons, is shown schematically in Figure~\ref{fig:SystemDesign}. The   ARNN carries out a Collective Classification of the compromised nodes (if any) for a 	$n$-node IP network denoted $V=\{v_1,~...~,v_n\}$.  For each network node $v_i$, the ARNN has two neurons $X_i$ and $Y_i$ that represent opposite views. $X_i$ indicates that $v_i$ is compromised, while $Y_i$ indicates that $v_i$ is not compromised.  Their corresponding numerical decision variables are $Q_i,~q_i\in [0,1]$, where $Q_i$ is the probability that $X_i$ is excited and $q_i$ is the probability that $Y_i$ is excited. $X_i$ has an excitatory connection $W^+_{ij}$ to every other neuron $X_j$ and an inhibitory connection $W^-_{ij}$ to all $Y_j$ neurons, and $Y_i$ has an excitatory connection $w^+_{ij}$ to every other neuron $Y_j$ and an inhibitory connection $w^-_{ij}$ to all $X_j$ neurons. None of the neuron can directly excite or inhibit themselves. Thus inside the  ARNN, the neurons of type $X$ excite other neurons of type $X$ and inhibit all neurons of type $Y$, and vice-versa for the neurons of type $Y$. The ARNN is ``self-critical'' in the sense that neurons of type $X$ try to supress the neurons of type $Y$, and vice-versa. $\Lambda_i$, is a non-negative real number that represents the output from the LAD  (local attack detector) at $v_i$ stating that $v_i$ has been compromised while $\lambda_i$ represents the LAD output at node $v_i$ stating that it has not been compromised. $\Lambda_i,~\lambda_i$ can be chosen from the corresponding probabilities outputted from the LADs acting as excitatory and inhibitory external input, respectively,  for each $X_i$, while they have the opposite effect as inhibitory and excitatory input for $Y_i$, respectively.

The two neurons  $X_i$  and $Y_i$ have  {\em internal states} $K_i(t)\geq 0$ and $k_i(t) \geq 0$, respectively.
If its internal state $K_i(t)$ is strictly positive, then the RNN neuron $X_i$ will fire spikes at exponentially distributed successive intervals, sending excitatory and/or inhibitory spikes at rates $W^+_{ij},~W^-_{ij}\geq 0$ to the other neurons in the ARNN. Similarly when $k_i(t)>0$ neuron $Y_i$ will fire spikes at rates and $w^+_{ij},~w^-_{ij}\geq 0$ for $Y_i$, respectively, to the other neurons $X_j$ and $Y_j$ in the ARNN. These firing rates are the ``weights'' that are learned with the training dataset using the algorithm described in \ref{Appendix}.

When any of the neurons $ \{X_i,~Y_i,~i=1,~...~ n\}$ receives an excitatory spike either from its external input or from another neuron, say at time $t$, its internal state will increase by $1$,
i.e. $K_i(t^+)=K_i(t)+1$ or $k_i(t^+)=k_i(t)+1$. Similarly if a neuron receives an inhibitory spike then its internal state decreases by $1$ provided it was previously at a positive state value, and its state does not change if it was previously at the zero value, i.e.  $K_i(t^+)=max[0,K_i(t)-1]$ or $k_i(t^+)=max[0,k_i(t)-1]$. 
Also  when a neuron fires, its internal state drops by $1$,
i.e. $K_i(t^+)=K_i(t)-1$ or $k_i(t^+)=k_i(t)-1$; note that a neuron can only fire if its state was previously positive.

We thus define the probability that these $2n$  neurons are ``excited'' or firing by:
\begin{eqnarray}
&&For~ X_i:~Q_i=\lim_{t\rightarrow\infty} Prob[K_i(t)>0],\\
&&For~Y_i:~q_i=\lim_{t\rightarrow\infty} Prob[k_i(t)>0],
\end{eqnarray}
and $Q_i$ is the variable that ``advocates'' that node $i$ is compromised, while the role of $q_i$ is to advocate the opposite.

Consider the following system of $2n$ equations for $Q_i,~q_i$, obtained from the RNN equations \cite{bib:Gelenbe1993}:
\begin{eqnarray} \label{RNN}
Q_i&=&\frac{\Lambda_i + \sum_{j=1}^n W^+_{ji}Q_j}{\lambda_i+\sum_{j=1}^n [W^+_{ij}+W^-_{ij}]+\sum_{j=1}^n w^-_{ji}q_j},\\
q_i&=&\frac{\lambda_i + \sum_{j=1}^n w^+_{ji}q_j}{\Lambda_i+\sum_{j=1}^n[w^+_{ij}+w^-_{ij}]+\sum_{j=1}^n W^-_{ji}Q_j},\nonumber
\end{eqnarray} 
where
\begin{equation}
W^+_{ii}=W^-_{ii}=w^+_{ii}=w^-_{ii}=0.
\end{equation}
Let $K(t)=(K_1(t),~...~,K_n(t))$ and $k(t)=(k_1(t),~...~,k_n(t))$, and define the vectors of non-negative integers $H=(H_1,~...~,H_n)$ and $h=(h_1,~...~,h_n)$.
From \cite{bib:Gelenbe1993}, we know that if the solution to the equations (\ref{RNN}) satisfy $0\leq Q_i,~q_i< 1$ for $1\leq i\leq n$, then the joint stationary distribution of the  ARNN's state is:
\begin{eqnarray} \label{prod}
&&\lim_{t\rightarrow\infty}Prob[K(t)=H,~k(t)=h~]\\ 
&&~~~~~~~~~~~=\prod_{i=1}^n Q_i^{H_i}(1-Q_i).q_i^{h_i}(1-q_i)~.\nonumber
\end{eqnarray}

\bigskip
 %More simply one can also use:
%\begin{equation}
%\Lambda_i=A^l_i,~\lambda_i=1-A^l_i,
%\end{equation}
%to estimate whether a Botnet attack is targeting
%node $i$ in Bucket $l$.
\noindent{\bf Note:} From  (\ref{prod}) we can see that if $Q_i>q_i$ then:
\begin{equation} \label{A}
\lim_{t\rightarrow \infty} Prob[K_i(t)>k_i(t)] = \frac{Q_i(1-q_i)}{q_i(1-Q_i)}>1.
\end{equation}

%Note that the outgoing firing rates of each of pair of neurons at a RNN node are such that:
%\begin{equation}
%If~ A(i,j)=0,~then~ W^+_{ij}=W^-_{ij}=w^+_{ij}=w^-_{ij}=0.
%\end{equation}
To simplify the learning algorithm, we restrict the weights in the following manner:
\begin{equation}
	W=W^+_{ij}+W^-_{ij} = w^+_{ij}+w^-_{ij},~i,j\in \{1,~..~n\},~i\neq j, \label{con}
\end{equation}
where $W>0$ is a constant representing the total firing or spiking rate any neuron $X_i$ or $Y_i$ 
towatds other neurons. 
This restriction also avoids having weights which take very large values. 
%$W^+_{ij}$ and $w^+_{ij}$ are initially set to specific values and then they will be updated using the learning algorithm described below.
%The weights in the equations for $Q_i$ and $q_i$ are also constrained in the following manner: 
%\begin{equation}
%	W^+_{ii}=w^+_{ii}=W^-_{ii}=w^-_{ii}=0,~ for ~any~ i,
%\end{equation}
%\begin{equation}
%W=W^+_{ij}+W^-_{ij} = w^+_{ij}+w^-_{ij},~\forall~i,j\in \{1,~..~n\},~i\neq j,
%\end{equation}
%where $W$ is a constant representing the total firing or spiking rate any neuron $X_i$ or $Y_i$ to all the other neurons with states$X_j,~Y_j$. 
We can write the $2n$ RNN equations (\ref{RNN}) as:
\begin{eqnarray} \label{Qq}
Q_i&=&\frac{\Lambda_i + \sum_{j=1}^n W^+_{ji}Q_j}{\lambda_i+(n-1)W+\sum_{j=1}^nw^-_{ji}q_j},\\
q_i&=&\frac{\Lambda_i + \sum_{j=1}^n w^+_{ji}q_j}{\Lambda_i+ (n-1)W+\sum_{j=1}^nW^-_{ji}Q_j}.\nonumber
\end{eqnarray}
On the other hand, the learning algorithm detailed in \ref{Appendix}
computes the values of $W^+_{ij},~w^+_{ij}$ for all the neuron pairs $i,j,~i\neq j$
so as to minimize an error based  cost function ${\bf E}$ using an appropriate training dataset
such as Kitsune \cite{mirsky2018kitsune,KitsuneKaggle}.

\section{Network Learning and Accuracy of Botnet Attack Prediction}\label{Experimental}

The data we use concerns the MIRAI Botnet Attack \cite{Soldatos2}. documented in the Kitsune dataset \cite{mirsky2018kitsune,KitsuneKaggle}
which contains  a total of $764,137$ individual
packets. The dataset contains $107$ network nodes identified by IP addresses,  and a given node may be both a source node for some packets, and a destination for other packets.

This publicly available dataset, which is already partially processed (by the providers of the dataset) contains the ground-truth that the providers held, regarding the packets which are
Botnet attack packets, and those which are not attack packets. Thus each packet is labeled as either an ``attack'' ($a=1$) or a ``normal'' packet ($a=0$), so that the Kitsune dataset
already contains the ``ground truth''. 
Since the dataset is quite large, some parts of the data may be used for training the attack detection algorithms, while other parts may be used for evaluating the effectiveness of them.

The data items in this dataset are the individual packets, where each packet can be denoted as 
$pk(t,s,d,a)$, where:
\begin{itemize}
	%\item  $I$ is its unique numerical identifier $1\leq I\leq I_M$, 
	\item $t$ is a time-stamp indicating when the packet is sent, 
	\item $s,d$ are the source and destination nodes of the packet,
	\item $a$ is the binary variable with $a=1$ for a packet that has been identified as an attack packet, and $a=0$ for a packet that has been identfied as a benign non-attack packet.
\end{itemize} 
It is interesting to note that this dataset is time varying. The obvious reason is that in the course of a Botnet attack the number of nodes that are compromised increases with the number of attacks which occur, and the number of attack packets obviously also increases as the number of compromised nodes increases. The Kitsune dataset does not incorporate the consequences of attack detection. Indeed if an attack is detected and the compromised nodes are progressively blacklisted, then the number of attack packets and the number of nodes that are compromised, may eventually decrease, but this is not incorporated in the Kitsune dataset.

Thus, since this data is based on an attack that is going unchecked, the initial part of the data contains hardly any attack packets, while the latter part contains many more attack packets, as would be expected. Whether a given node is compromised or not also depends on the amount of traffic it receives from compromised nodes, as this traffic may contain attack packets capable of compromising the destination node.
Thus detecting whether a network node is compromised or not, does not only depend on its own behaviour, i.e.
on whether it sends attack packets, but also on whether it has received traffic from other compromised nodes.

\subsection{Processing the MIRAI Botnet Data}

These $764,137$ packets in \cite{KitsuneKaggle} cover a consecutive time period of roughly $7137$ seconds (approximately $2$ hours). Thus we aggregate
the data in a more compact form by grouping packets into successive time $10$ second 
time slots whose length is denoted by $\tau$.
The choice of $\tau=10~secs$  is based on the need to have a significant number of $\approx 713$ time slots, and to have a statistically significant number of packets in each slot. Since we have $107$ nodes, the average number of packets per node in each slot is also approximately $10$.

The packets within each successive slot are thus
grouped into ``buckets'', where $B^l$ denotes the $l-th$ bucket, i.e.
the set of packets whose time stamp lies between $(l-1)\tau$ and $l\tau$ seconds:
\begin{equation}
B^l = \{pk(t,s,d,a),~(l-1)\tau\leq t <l \tau\},~\tau=10~secs.\nonumber
\end{equation}
%Within bucket $B^l$, we denote the set of packets which are {\em sent} from node $s$ to any of the destinations $d$ as: \begin{equation} S^l(s)= \{pk(I,t,s,d,a),~\forall d,~\forall a,~(l-1)\leq t <l \},\nonumber \end{equation} and the set of packets which are {\em received} at node $d$ from all sources $s$ in $B^l$ are denoted: \begin{equation} R^l(d)= \{pk(I,t,s,d,a),~\forall s,~\forall a,~(l-1)\leq t <l \}.\nonumber \end{equation} 

Let $S^l(s)$ denote the set of packets that have been transmitted by node $s$ {\em until the end of the $l-th$ time slot}:
\begin{equation}
S^l(s)= \{pk(t,s,d,a),~\forall d,~\forall a,~0\leq t <l \tau \},
\end{equation} 
and, let $R^l(d)$ denote the set of packets that have been received by node $d$ in the same time frame: 
\begin{equation}
R^l(d)= \{pk(t,s,d,a),~\forall s,~\forall a,~0\leq t <l \tau\}~.
\end{equation} 
Then $A^l_d$ is the {\em attack ratio} which represents the ratio of attack packets, among all packets received by node $d$ at the end of $l-th$ slot and is computed as
\begin{eqnarray}
&&If~~| R^l(d)| >0:\nonumber\\
&&A^l_d=\frac{|\{pk(t,s,d,1),~\forall s,~0\leq t <l\tau\}| }{| R^l(d)| },\\
&&Else~~A^l_d=0\nonumber,
\end{eqnarray} 
while $K^l_s$ is the {\em proportion of compromised packets} which is the ratio of attack packets sent by node $s$ at the end of the same slot, given by:
\begin{eqnarray}
&&If~~| S^l(s)| >0:\nonumber\\
&&K^l_s=\frac{|\{pk(t,s,d,1),~\forall d,~0\leq t <l\tau\}| }{| S^l(s)| },\\
&&Else~~K^l_s=0. \nonumber
\end{eqnarray} 
Since any node $i$ may be a source or destination, or both a source and destination, of packets, $A^l_i$ and $K^l_i$ are, respectively, the input and output ground truth data regarding which nodes are attacked, and which nodes are compromised at the end of $l-th$ time slot.

In addition, for each node $i$, we define the binary variable regarding the \emph{ground truth}, denoted by $G^l_i$ as:
\begin{equation}\label{groundtruth_binary_decision}
G^l_i = \mathbf{1}\left[K^l_i > \Theta\right],
\end{equation}
where $\mathbf{1}\left[L\right]=1$ if $L$ is true and $0$ otherwise, where $\Theta\in[0,1]$ is a threshold. Thus, at the end of the $l-th$ slot, if $G^l_i=1$ the ground truth indicates that node $i$ 
has been compromised. If $G^l_i=0$ then node $i$ is considered not to be
compromised.

\subsection{The  ARNN Error Functio $E$} \label{Learn}

Let us call ''{\bf TrainData}'' the subset of time slots
used for Training the  ARNN. The manner in which this subset is selected from the MIRAI dataset is detailed below. 
Since we wish to predict whether each of the $n$ nodes has been compromised given the data about attacks, the error function to be minimized by the learning algorithm takes the form:
\begin{eqnarray}
{\bf E}&=&\frac{1}{2}\sum_{l\in {\bf TrainData}}\sum_{i=1}^n \big[\big(Q^l_i(A^l_i)-K^l_i\big)^2 \nonumber\\ 
&&~~~~~~~~~+\big(q^l_i(1-A^l_i)-(1-K^l_i)\big)^2 \big], \label{cost}
\end{eqnarray}
where the functions $Q^l_i(A^l_i)$ and $q^l_i(1-A^l_i)$ are computed by the ARNN using equation (\ref{Qq}) as follows:
\begin{eqnarray} 
&&Q^l_i(A^l_i)=\nonumber\\
&&\frac{A^l_i + \sum_{j=1}^n W^+_{ji}Q^l_j(A^l_i)}{(1-A^l_i)+ (n-1)W+\sum_{j=1}^nw^-_{ji}q^l_j(1-A^l_i)},\nonumber
\end{eqnarray}

\begin{eqnarray}
q^l_i(1-A^l_i)&&=\nonumber\\
&&\frac{(1-A^l_i) + \sum_{j=1}^n w^+_{ji}q^l_j(1-A^l_i)}{A^l_i + (n-1)W+\sum_{j=1}^nW^-_{ji}Q^l_j(A^l_i)}.\nonumber
\end{eqnarray}

For each node $i$, we define the {\bf binary decision} of the output of the  ARNN, denoted by the binary variable $Z_i$ as 
\begin{equation}\label{binary_decision}
Z^l_i = \mathbf{1}\left[L^l_i = \frac{Q^l_i(1-q^l_i)}{q^l_i(1-Q^l_i)}>\gamma\right],
\end{equation}
where $\gamma\in[0,\infty]$ is a ``decision threshold''. Thus, at $l-th$ slot, if $Z^l_i=1$ the  ARNN indicates that node $i$ has been compromised, while if $Z^l_i=0$ then ARNN considers that node $i$ is not compromised.

Then, we perform two distinct experiments:

\subsubsection{Experiment I: Offline Training of  ARNN}

To construct a balanced training dataset  {\bf TrainData}  for the  ARNN, the sequence of slots was scanned chronologically from the beginning of the whole MIRAI dataset until the first slot was found that contained some  nodes that had been compromised. Specifically, this was in $l^*-th$ slot with $l^*=445$ in the MIRAI dataset.  

Then, the training set ${\bf TrainData}$ with a total of $25$ time slots was constructed as follows:
	\begin{eqnarray}
	&&	{\bf TrainData}=\nonumber\\
	&&	\{(A^l_i,K^l_i),~l=l^*-12,...,{l^*+12};~i=1,...,n\},\nonumber \label{CTS}
	\end{eqnarray}
	of which the first $12$ have very few attack packets, while the following $13$ all contain a significant number of attack packets. 
	
The test set, denoted by ${\bf TestData}$, is composed of {\em all the remaining} time slots which have not used for training the  ARNN:
	\begin{eqnarray}
	&&{\bf TestData}=\nonumber\\
	&&\{(A^l_i,K^l_i),~l=\{1,..., l^*-13\} \cup \{l^*+13,...,713\};\nonumber\\
	&&~~~~~~~~~~~~~~i=1,...,n\},\nonumber 
	\end{eqnarray}

\subsubsection{Experiment II: Online (Incremental) Training  of  ARNN}

In this part, ARNN's training took place online, along with testing, which represents the case where there is no available training set offline. To this end, it was used for prediction on every slot $l$ and also if $mod(l, 6)=0$ it was trained at the end of slot $l$. That is, we perform testing for $10$ second slots and training for $1$ minute slots. 

Accordingly, on each ``training slot'' $l$ for which $mod(l, 6)=0$, the training set ${\bf TrainData}$ for incremental learning was constructed as follows:
	\begin{equation*}
	{\bf TrainData}= \{(A^{l'}_i,K^{l'}_i),~{l'}=l-5,...,{l};~i=1,...,n\}.
	\end{equation*}
Recall that {\bf TrainData} is updated for each $l$ such that $mod(l, 6)=0$, so that the  ARNN's weights ($W^+_{ij}$ and $w^+_{ij}$) are updated based on {\bf TrainData} at the end of slot $l$, without reinitializing the weights. 

\subsection{Other Machine Learning Models  Used for Comparison}\label{sec:ML}

\begin{figure}[h!]
	\centering 

	\includegraphics[scale=0.55]{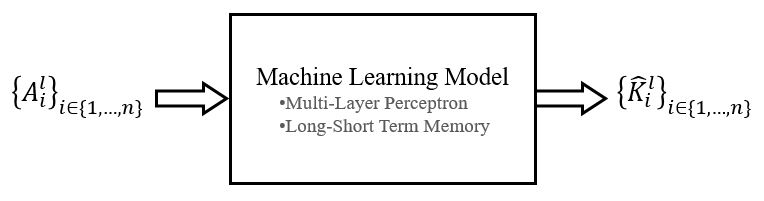}
	\caption{High-level architecture that shows the inputs and outputs at each slot $l$ for the ML techniques that are used in the comparison with  the ARNN, where $\hat{K}_i^l$ denotes the predicted compromised ratio of IP Address $i$ at slot $l$ by any considered ML model.
	}
	\label{fig:ML_arch}
\end{figure}

For both Experiments I and II, the performance of  the ARNN is also compared with those obtained with two well-known ML models: the Multi-Layer Perceptron (MLP) and the Long-Short Term Memory (LSTM) neural network. We now briefly present the specific architectures of these models which we use during our experimental work, and Figure~\ref{fig:ML_arch} displays the inputs and outputs which are common to the ML models.  

Then, based on these input-output sets, each ML model is used as follows:
\begin{itemize}
	
	%\item \textbf{Linear Regression} is considered as the linear baseline model. 
		
	%\item \textbf{KNN} is implemented using the \emph{scikit-learn} library \cite{scikit-learn} in Python. The number of neighbours in the KNN is set to the number of slots in the {\bf TrainData}, which equals $25$ for Experiment I and $6$ for Experiment II.
	
	\item \textbf{MLP}, which is a feedforward (fully-connected) neural network, is comprised of three hidden layers and an output layer, where there are $n$ neurons at each layer. A sigmoidal activation function is used for each neuron in the network. 
	
	\item \textbf{LSTM}, which is a recurrent neural network, is comprised of an lstm layer, two hidden layers and an output layer, where there are $n$ lstm units or neurons at each layer. A sigmoidal activation function is used for each neuron in the network. 
	
\end{itemize}

\section{Experimental Results}\label{Eval}

We now evaluate the performance of  the ARNN model and compare it with the performance of some existing techniques for Experiment I and Experiment II, respectively. Note that we set the learning rate $\eta=0.1$ in the algorithm of \ref{Appendix}.

\subsection{Experiment I - Offline Training of  the ARNN} 

\begin{figure}[h!]
	\centering 
	\includegraphics[scale=0.25]{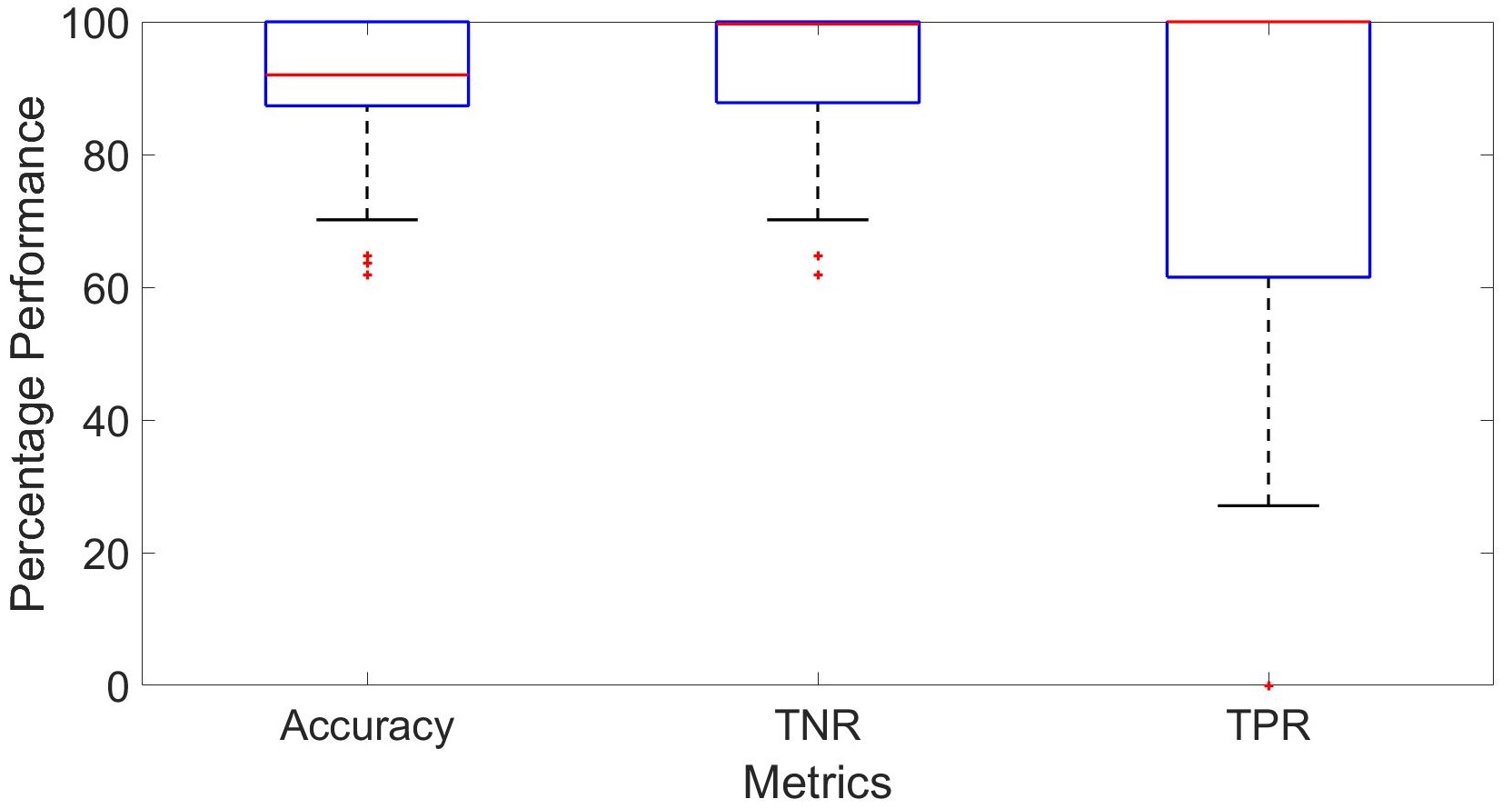}
	\caption{Box-plots of the Accuracy, TNR and TPR performance of  ARNN over IP Addresses, where each of box-plot shows the calculated statistics (e.g. median) based on the results presented in Figures~\ref{fig:Accuracy_offline},~\ref{fig:TNR_offline},~and~\ref{fig:TPR_offline}, respectively
	}
	\label{fig:BoxPlot_offline}
\end{figure}

\begin{figure}[h!]
	\centering 
	\includegraphics[scale=0.25]{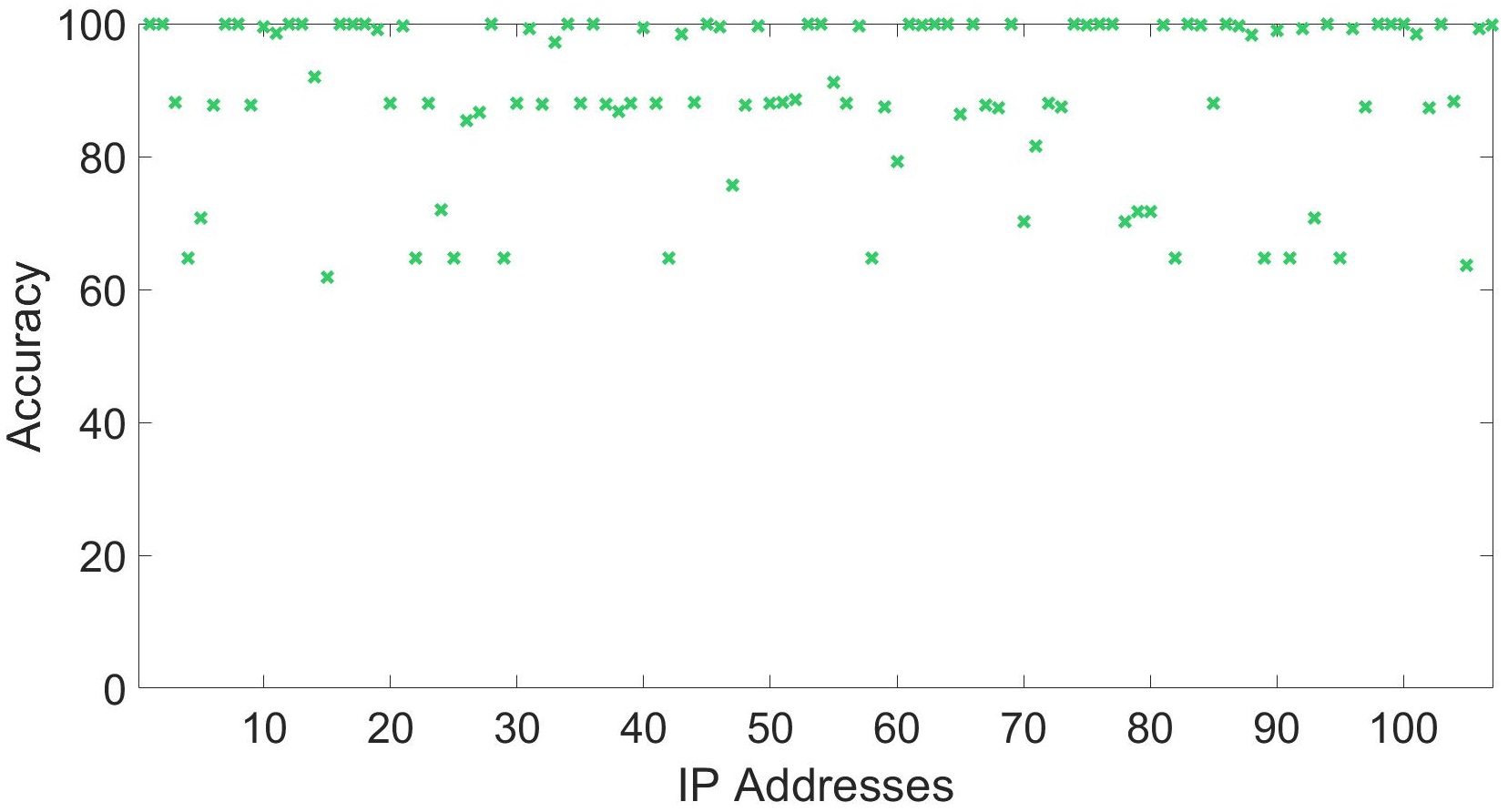}
	\caption{Evaluation of the average accuracy over all packets of each IP Address $i \in \{1, \dots, 107\}$ in ${\bf TestData}$. The accuracy is computed by comparing the binary decision in the ground truth $G_i^l$ and the binary decision of ARNN $Z_i^l$.% with $\Theta=0.2$ and $\gamma=0.1$. %We observe that the accuracy equals $100\%$ for $78\%$ of all IP Addresses while it is below $80\%$ for only $7$ IP Addresses.
		%time slots $l \in \{1, \dots, 713\}\backslash \{433, \dots, 458\}$, where $\gamma = 0.16$. Note that if $G_i^l \neq 1$ $\forall l \in \{1, \dots, 713\}\backslash \{433, \dots, 458\}$, then TPR does not exist for IP address $i$.
	}
	\label{fig:Accuracy_offline}
\end{figure}

\begin{figure}[h!]
	\centering 
	\includegraphics[scale=0.25]{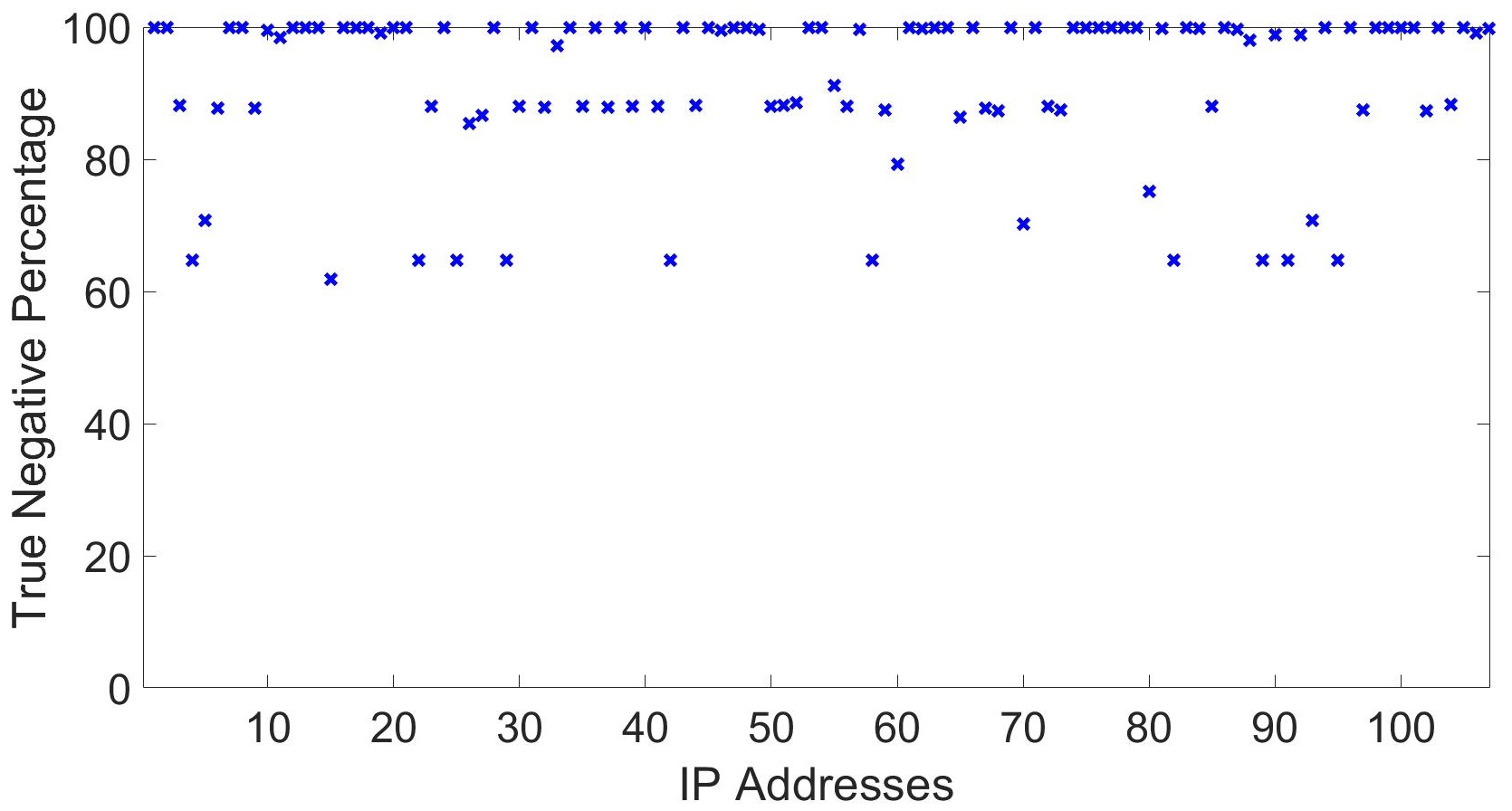}
	\caption{Evaluation of the average percentage TNR over all packets of each IP Address $i \in \{1, \dots, 107\}$ in ${\bf TestData}$. For each $i$, TNR is computed by comparing $G_i^l$ and $Z_i^l$ for the values of $l$ where $G_i^l=0$. 
	}
	\label{fig:TNR_offline}
\end{figure}

\begin{figure}[h!]
	\centering 
	\includegraphics[scale=0.25]{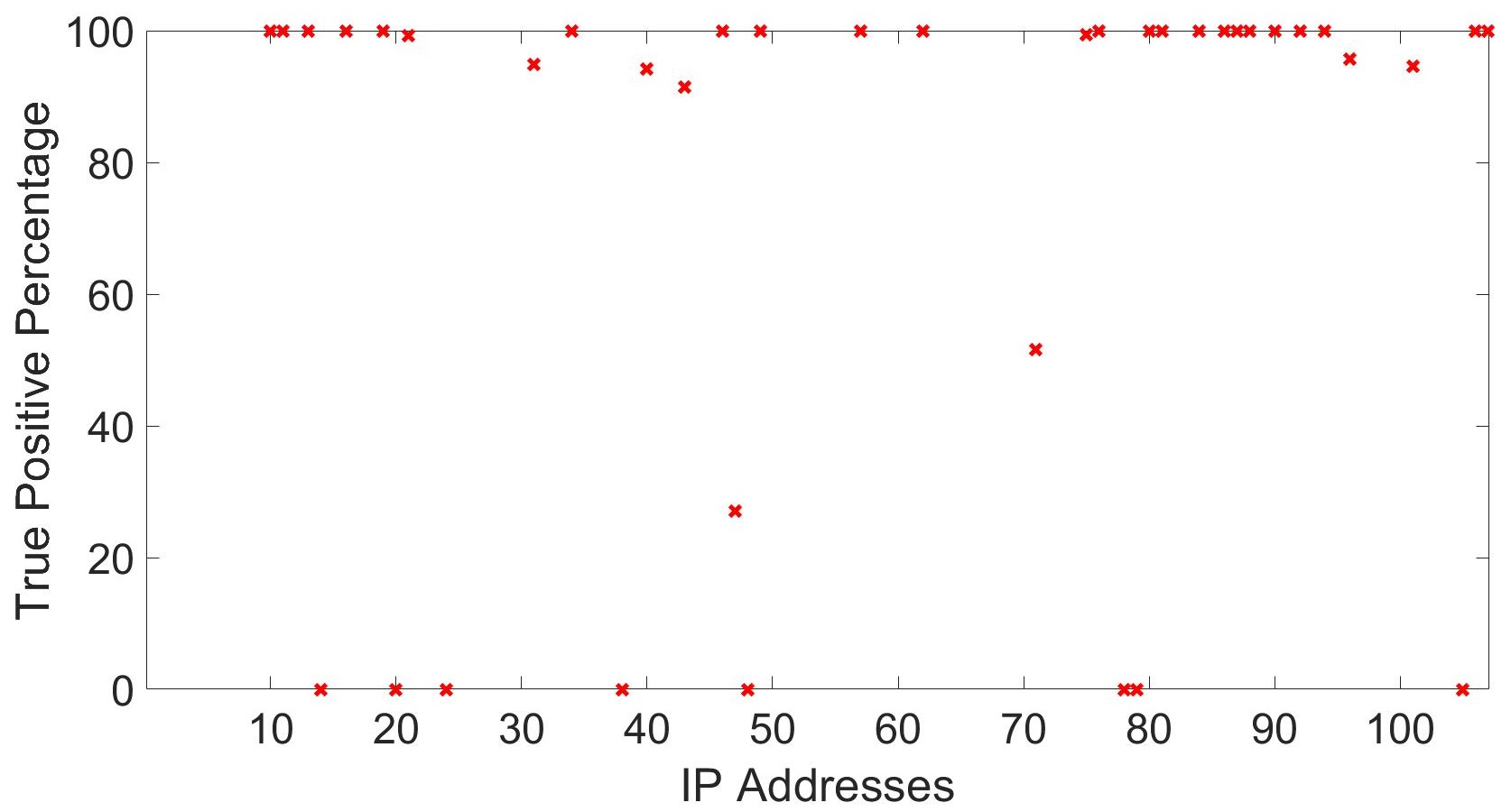}
	\caption{Evaluation of the average percentage TPR over all packets of each IP Address $i \in \{1, \dots, 107\}$ in ${\bf TestData}$. For each $i$, TPR is computed by comparing $G_i^l$ and $Z_i^l$ for the values of $l$ where $G_i^l=1$. Note that if $G_i^l=0$ for an IP Address $i$ for any $l$ in ${\bf TestData}$ (that is, the ground truth indicates that IP Address $i$ has not been compromised within the observation period of the dataset), TPR does not exist for $i$. Accordingly, in the considered dataset TPR exists for $39$ IP Address. 
	}
	\label{fig:TPR_offline}
\end{figure}

We set $\Theta=0.3$ and $0.96 \leq \gamma \leq 1$, and summarize the statistics of Accuracy, True Negative Rate (TNR) and True Positive Rate (TNR) performances of  ARNN, which are presented in detail in Figures~\ref{fig:Accuracy_offline},~\ref{fig:TNR_offline},~and~\ref{fig:TPR_offline}, respectively. Figure~\ref{fig:BoxPlot_offline} displays a box-plot that shows the statistics over all the IP Addresses. These results show that  ARNN achieves a high performance with very few outliers in regards of Accuracy, TNR, and TPR. The median accuracy is about $92\%$ while the first quartile is at $87\%$; that is, accuracy is above $87\%$ for $75\%$ of IP addresses. The median of TNR is almost $100\%$; that is, there are almost no false alarms (TNR$=100\%$) for more than $50\%$ of IP addresses. Also, the median of TPR equals $100\%$ and the first quartile is about $62\%$. Thus the TPR equals $100\%$ for more than $50\%$ of IP addresses while it is lower than $62\%$ for only less than $25\%$ of addresses.

Figure~\ref{fig:Accuracy_offline} displays the average decision accuracy for each IP Address $i \in \{1, \dots, 107\}$. The results in this figure show that the accuracy of  ARNN is above $95\%$ for $50\%$ of the IP Addresses while it is between $62\%$ and $80\%$ for only $20\%$ of addresses and does not decrease below $62\%$. Next, Figure~\ref{fig:TNR_offline} presents average percentage TNR of  ARNN for each IP Address. The results in this figure show that TNR is above $95\%$ for $59\%$ of all IP Addresses, and it is between $62\%$ and $80\%$ for $15\%$ of addresses. Lastly, Figure~\ref{fig:TPR_offline} displays the percentage average TPR for $39$ IP Addresses for which are considered compromised at least once in the ground truth. The results in this figure show that TPR is greater than $95\%$ for $64\%$ of IP addresses while it is above $90\%$ for more than $74\%$ of the addresses.

%Lastly, in order to observe the prediction ability of the  ARNN, we analyze the distribution of Likelihood Ratio (LR) predicted by  ARNN and compare that with LR in Ground Truth. This comparison is presented in Fig.~\ref{fig:Hist_offline}, where the results show that the distribution of the LR that is predicted by  ARNN is highly similar to that is computed with the Ground Truth. In addition, based on the results in this figure, we measured that the difference between Ground Truth and  ARNN is $136$ for the total number of occurrences of LRs that are greater than $1$ (or, symmetrically, that are less than $1$).

%\begin{figure}[h!]
%	\centering 
%	\includegraphics[scale=0.32]{Hist_25samplesTraining.jpg}
%	\caption{Comparison of the histograms of Likelihood Ratio that is predicted by  ARNN and that is computed with the Ground Truth in  ${\bf TestData}$. For the better visualization, the Likelihood Ratios are saturated at $3$. }
%	\label{fig:Hist_offline}
%\end{figure}

\subsection{Online (Incremental) Training  of  the ARNN} 

\begin{figure}[h!]
	\centering 
	\includegraphics[scale=0.25]{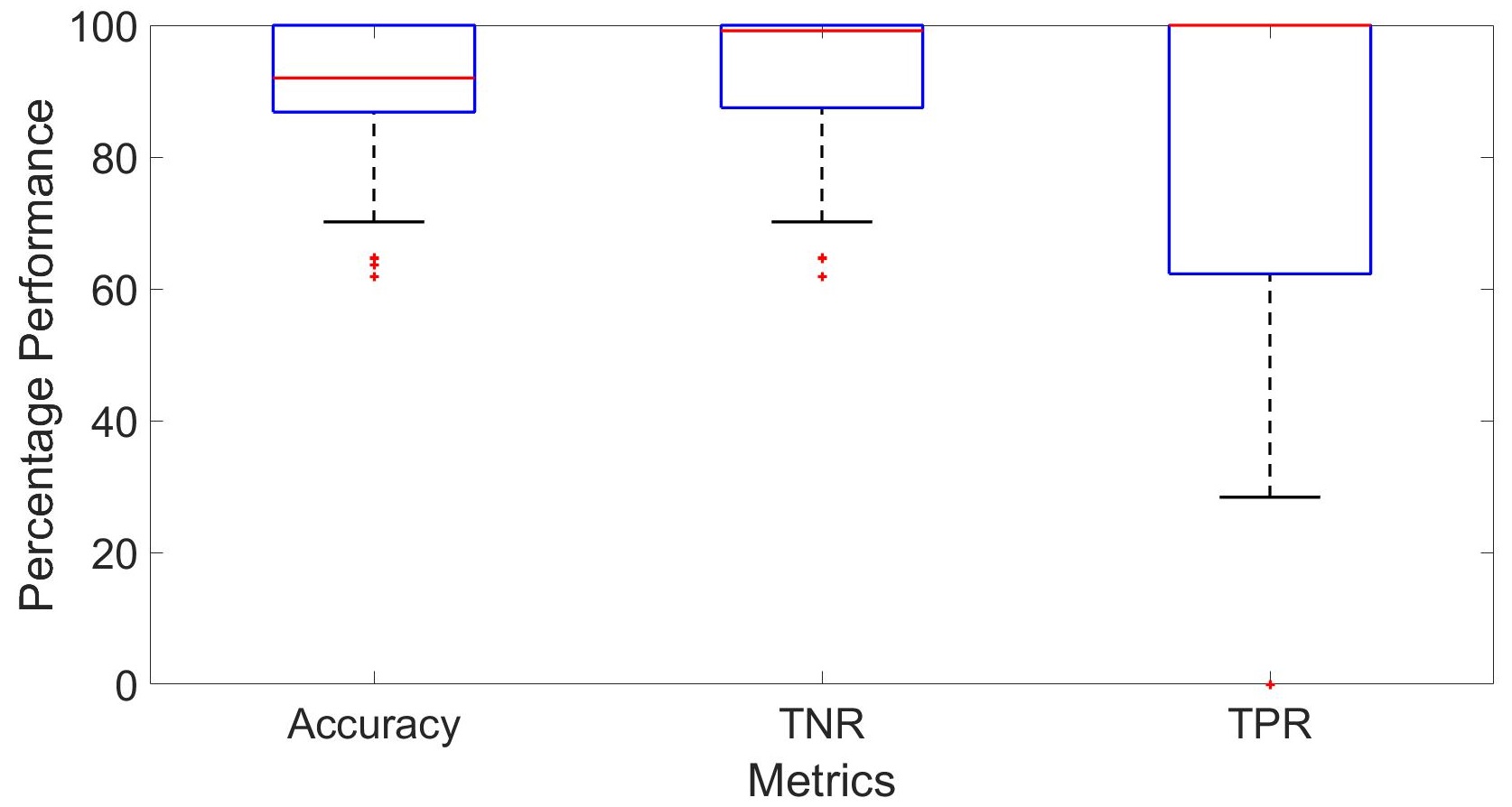}
	\caption{Box-plots of the Accuracy, TNR and TPR performance of  ARNN over all IP Addresses, where each box-plot shows the calculated statistics (e.g. median) based on the results presented in Figures~\ref{fig:Accuracy_online},~\ref{fig:TNR_online},~and~\ref{fig:TPR_online}, respectively
	}
	\label{fig:BoxPlot_online}
\end{figure}

\begin{figure}[h!]
	\centering 
	\includegraphics[scale=0.25]{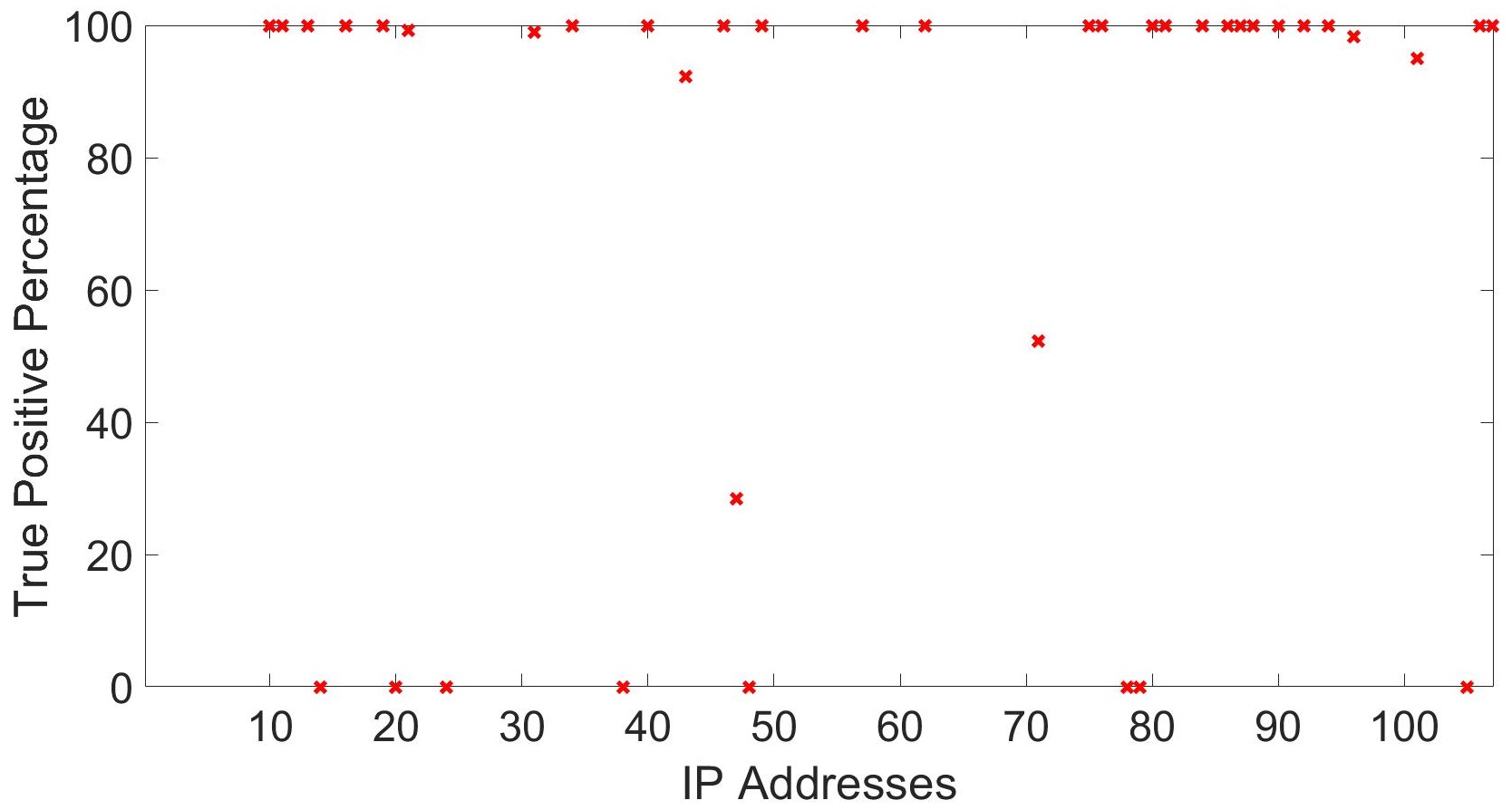}
	\caption{Evaluation of the average percentage TPR over all packets of each IP Address $i \in \{1, \dots, 107\}$ in ${\bf TestData}$. For each $i$, TPR is computed by comparing $G_i^l$ and $Z_i^l$ for the values of $l$ where $G_i^l=1$. Note that if $G_i^l=0$ for an IP Address $i$ for any $l$ (that is, the ground truth indicates that IP Address $i$ has not been compromised within the observation period of the dataset), TPR does not exist for $i$. Accordingly, in the considered dataset TPR exists for $39$ IP Address. %In addition, we set $\Theta=0.5$ and $\gamma=0.16$. We observe that  ARNN either identifies the compromised IP Address with $100\%$ success over time or cannot identify it at all. The sparse varying TPR performance among IPs may be due to the offline training and the selection of {\bf TrainData}.
	}
	\label{fig:TPR_online}
\end{figure}

Having set $\Theta=0.3$ and $0.96 \leq \gamma \leq 1$, we obtain the Accuracy, TNR, and TPR of  ARNN with online training shown in Figure~\ref{fig:TPR_online} in the form of box-plots. In this figure, we see that median accuracy equals $92\%$ while the first quartile equals $87\%$. That is, the accuracy is above $87\%$ for $75\%$ of all IP Addresses. The TNR is above $99\%$ for at least $50\%$ of IP Addresses. The median TPR equals $100\%$; that is, at least $50\%$ (exactly $62\%$) of IP Addresses are $100\%$ accurately identified as compromised. When the results in Experiment II in Figure~\ref{fig:BoxPlot_online} are compared with those of Experiment I of Figure~\ref{fig:BoxPlot_offline}, we see that TPR increases slightly with online training and TNR remains almost the same. In addition, recall that online training is simpler since it does not require data collection, as offline training does.

\begin{figure}[h!]
	\centering 
	\includegraphics[scale=0.25]{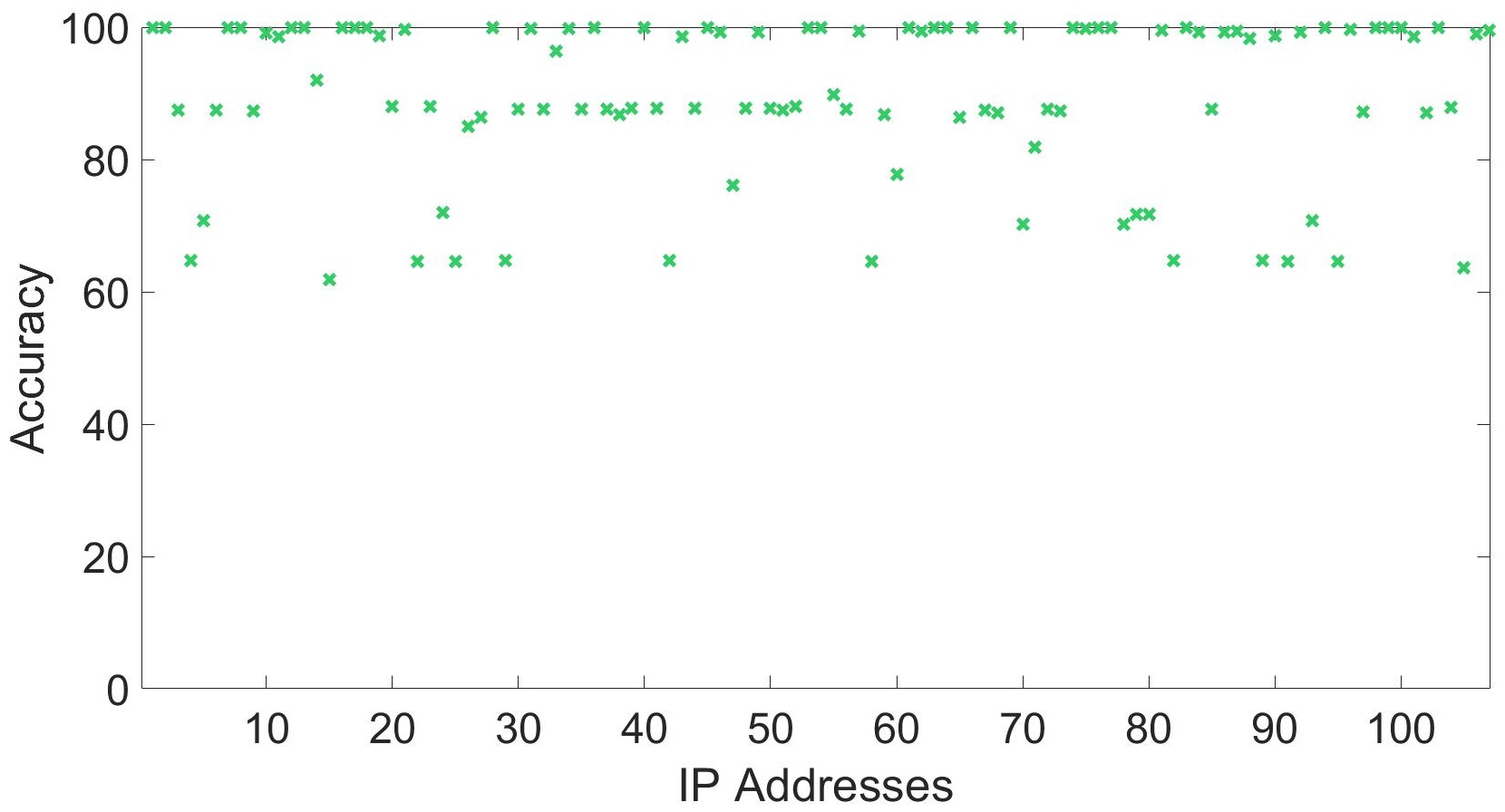}
	\caption{Evaluation of the average accuracy over all packets of each IP Address $i \in \{1, \dots, 107\}$. The accuracy is computed by comparing the binary decision in the ground truth $G_i^l$ and the binary decision of ARNN $Z_i^l$.
	}
	\label{fig:Accuracy_online}
\end{figure}

\begin{figure}[h!]
	\centering 
	\includegraphics[scale=0.25]{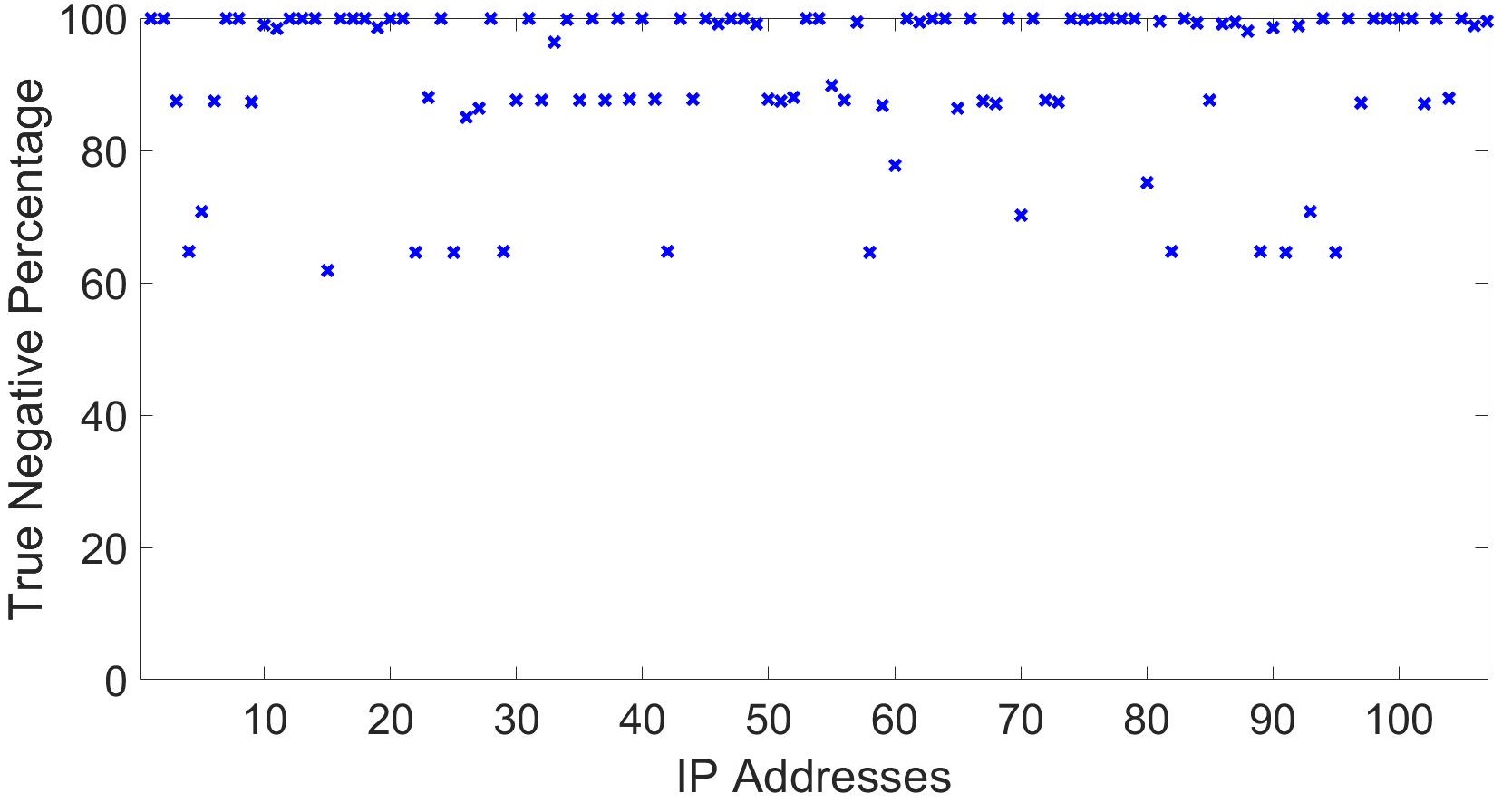}
	\caption{Evaluation of the average percentage TNR over all packets of each IP Address $i \in \{1, \dots, 107\}$. For each $i$, TNR is computed by comparing $G_i^l$ and $Z_i^l$ for the values of $l$ where $G_i^l=0$. 
	}
	\label{fig:TNR_online}
\end{figure}

Figure~\ref{fig:Accuracy_online} presents the average accuracy of  ARNN for each IP Address $i$ is displayed. The results in this figure show that the accuracy of  ARNN is above $95\%$ for $50\%$ of IP Addresses while it is between $62\%$ and $80\%$ for only $20\%$ of addresses and does not decrease below $62\%$. Next, we present the average percentage TNR in Figure~\ref{fig:TNR_online}, and show that the TNR is above $95\%$ for $59\%$ of IP Addresses. Moreover, Figure~\ref{fig:TPR_online} displays the average percentage TPR for individual IP Addresses, where for IP Address $i$, TPR is presented only if $G_i^l=1$ for at least a single value of $l$. The results in this figure show that percentage TPR is greater than $95\%$ for $72\%$ of IP Addresses, while TPR under offline training is shown (in Fig.~\ref{fig:TPR_offline}) to be above $95\%$ for $64\%$ of IP Addresses. Hence, one may observe that  ARNN achieves significantly higher TPR when it is trained online.

\subsection{Performance Comparison}

We now compare the performance of  ARNN with that of MLP and LSTM neural networks with respect to the mean of each Accuracy, TNR, TPR, and F1 Score. %The F1 Score is the harmonic mean of the sensitivity and precision of the attack decisions. It takes values in the range [0, 1], where 1 indicates perfect sensitivity and precision, and 0 indicates vice-versa.
The traditional F-measure or $F_1$ score is computed as
\begin{equation}
F_1=2\frac{Precision.Recall}{Precision+Recall}=\frac{TP}{TP+\frac{1}{2}(FP+FN)},
\end{equation}

\begin{figure*}[t!]
	\centering 
	\includegraphics[width=0.475\textwidth]{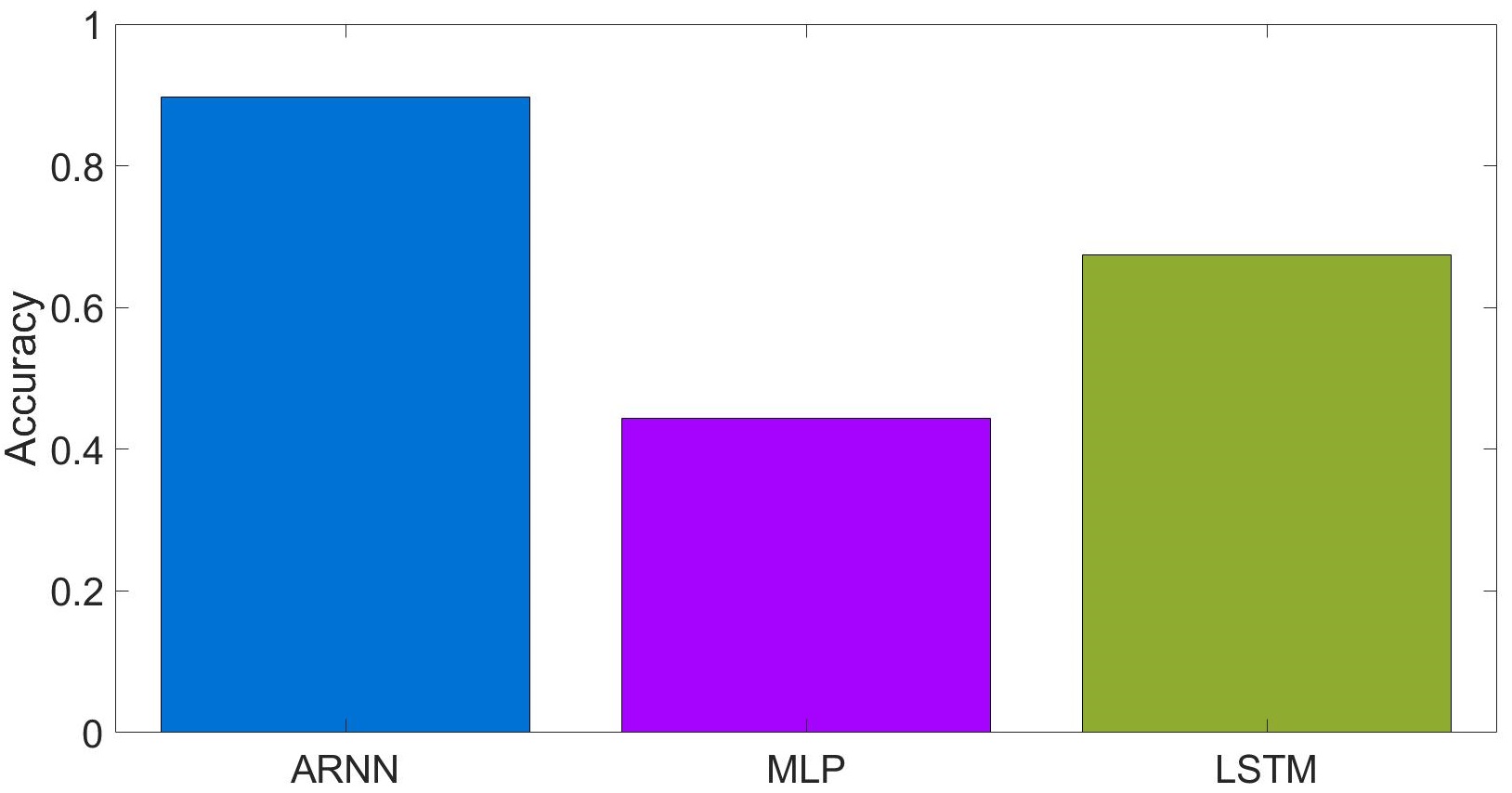}\hfill \includegraphics[width=0.475\textwidth]{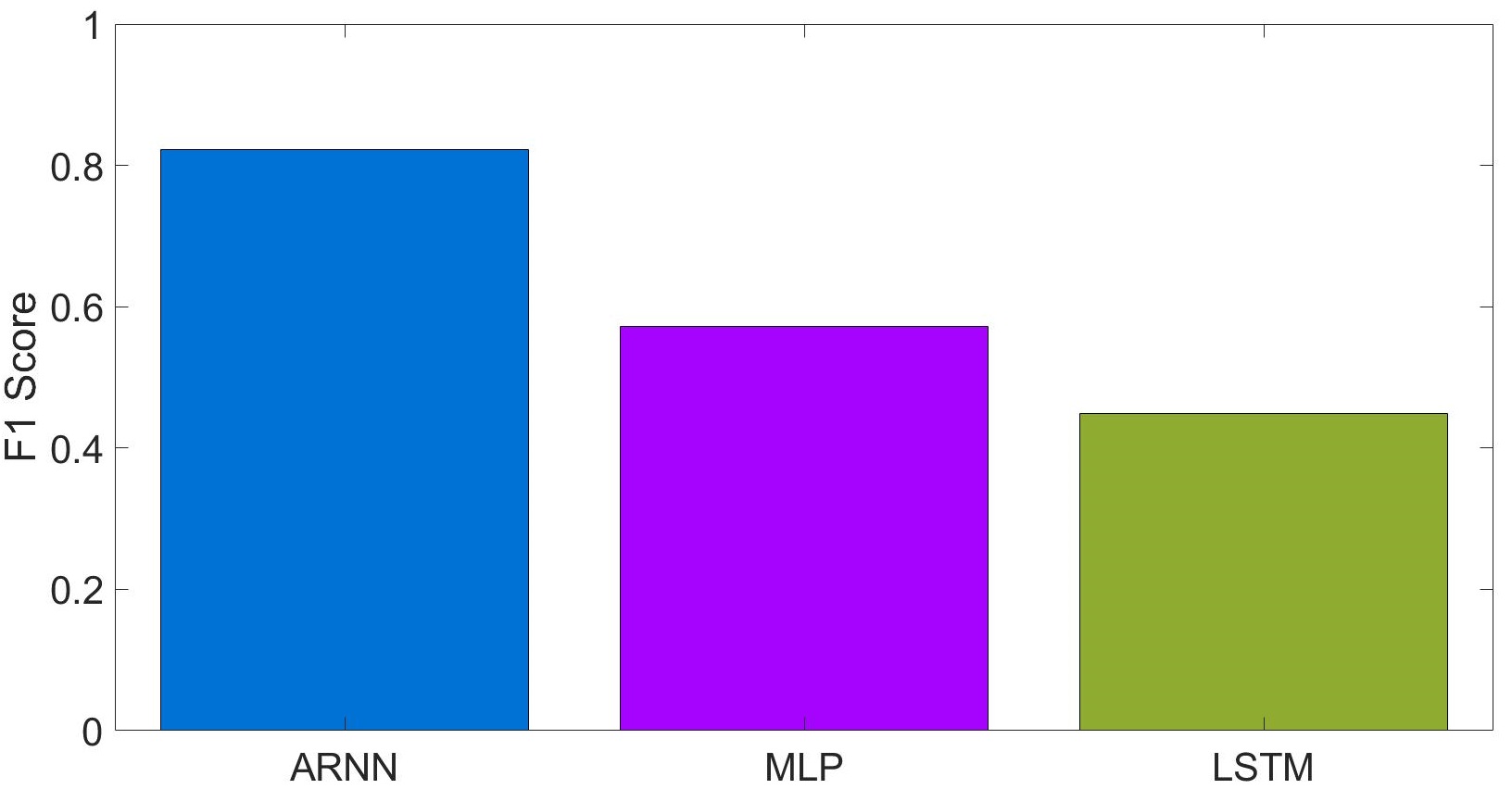}
	
	\vspace{0.5cm}
	
	\includegraphics[width=0.475\textwidth]{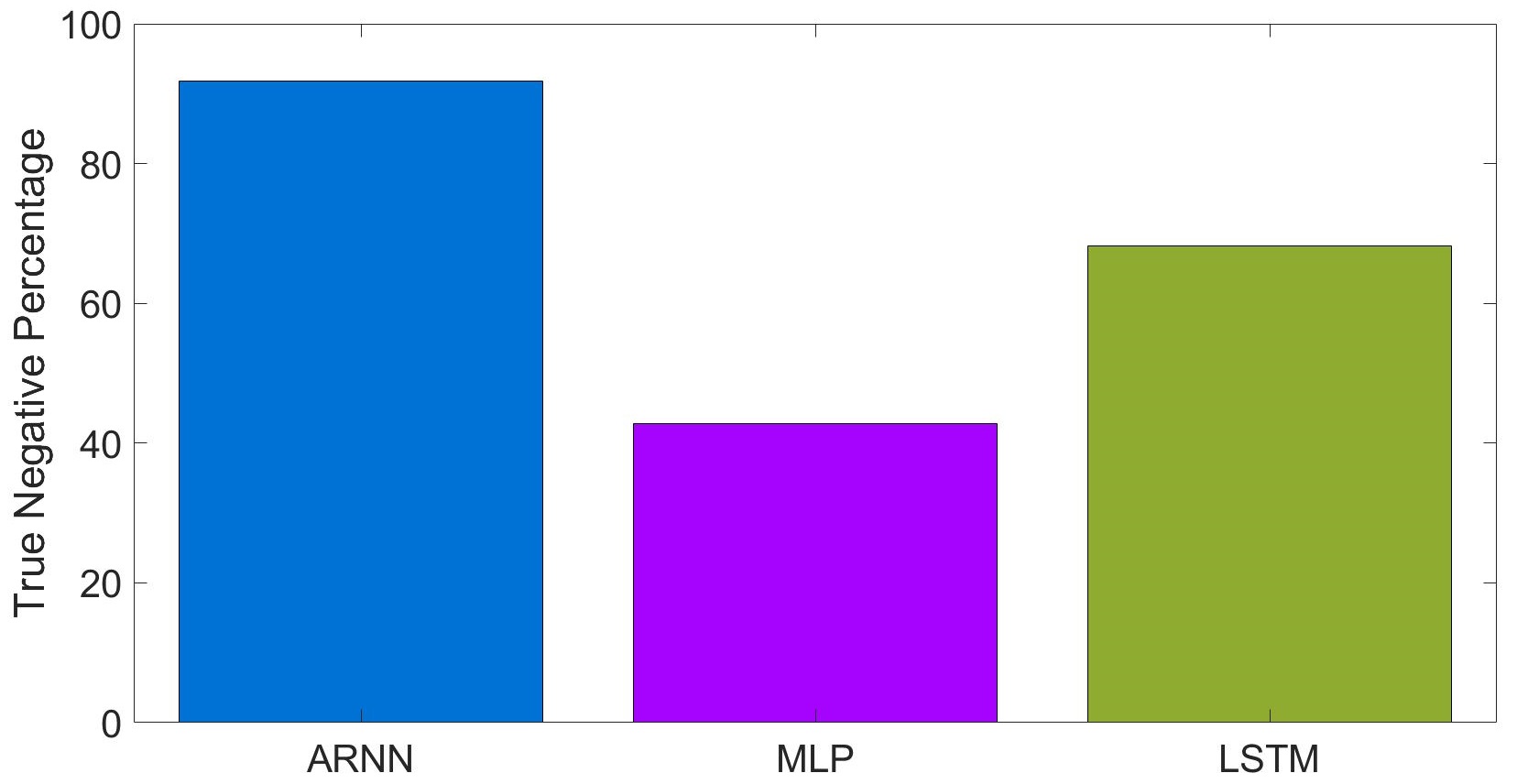}\hfill \includegraphics[width=0.475\textwidth]{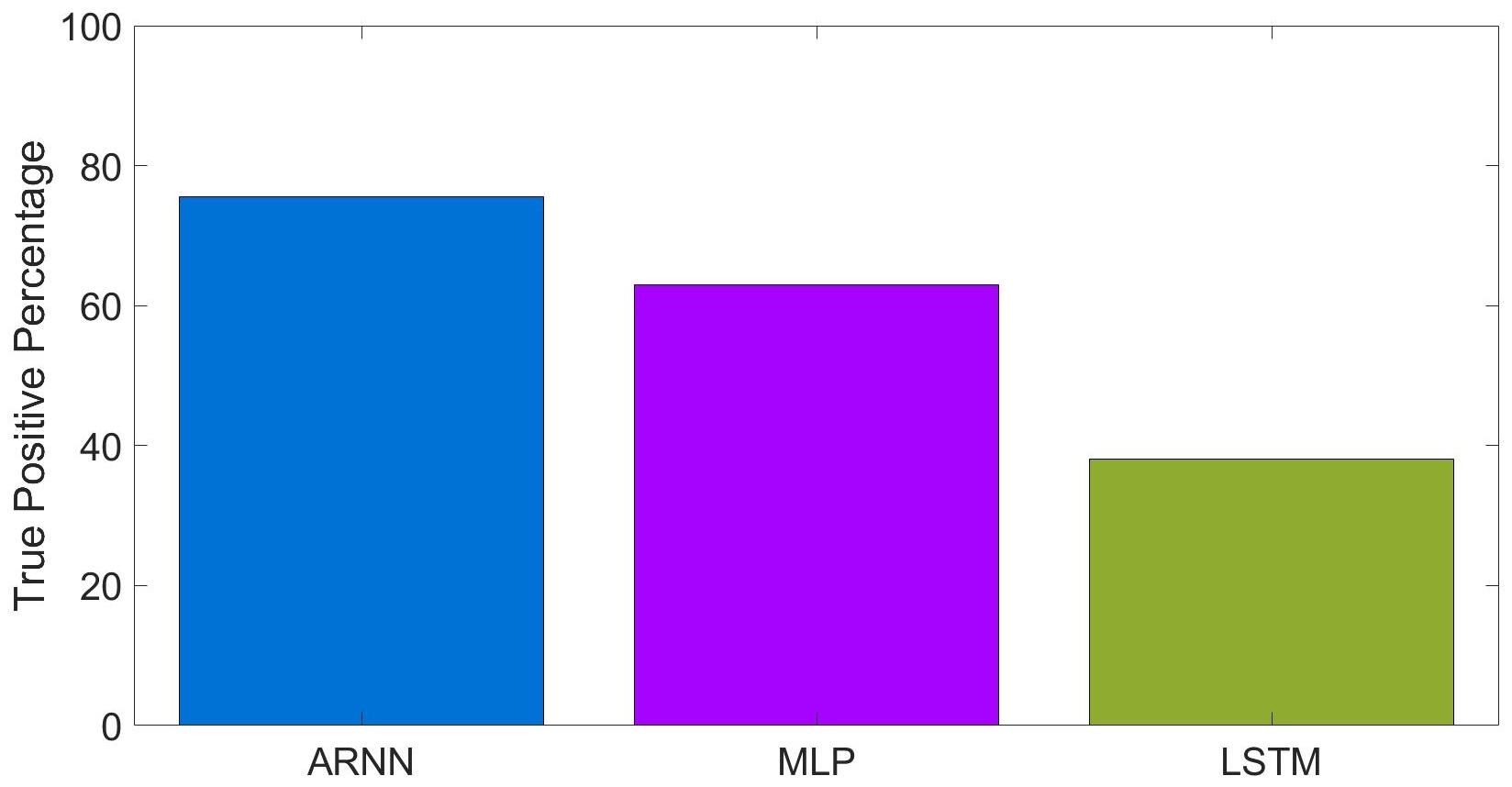}
	\caption{Performance comparison between  ARNN, MLP and LSTM for Experiment I (where each model is trained offline) with respect to \textbf{(top left)} Accuracy, \textbf{(top right)} F1 Score, \textbf{(bottom left)} percentage TNR, and \textbf{(bottom right)} percentage TPR}
	\label{fig:Comparison_offline}
\end{figure*}

%\subsubsection{Experiment I - Offline Training} 
First, Figure~\ref{fig:Comparison_offline} presents the performance comparison of neural network models for Experiment I (offline training), where the results show that the  ARNN model significantly outperforms all of the other techniques with respect to all Accuracy, F1 Score, TNR, and TPR. In addition, we also see that LSTM is more successful than MLP for identifying uncompromised nodes (Figure~\ref{fig:Comparison_offline} (bottom left)) while MLP identifies the compromised nodes more successfully than LSTM (Figure~\ref{fig:Comparison_offline} (bottom right)). However,  ARNN outperforms LSTM by $24\%$ with respect to TNR and MLP by $13\%$ with respect to TPR. 

\begin{figure*}[h!]
	\centering 
	\includegraphics[width=0.475\textwidth]{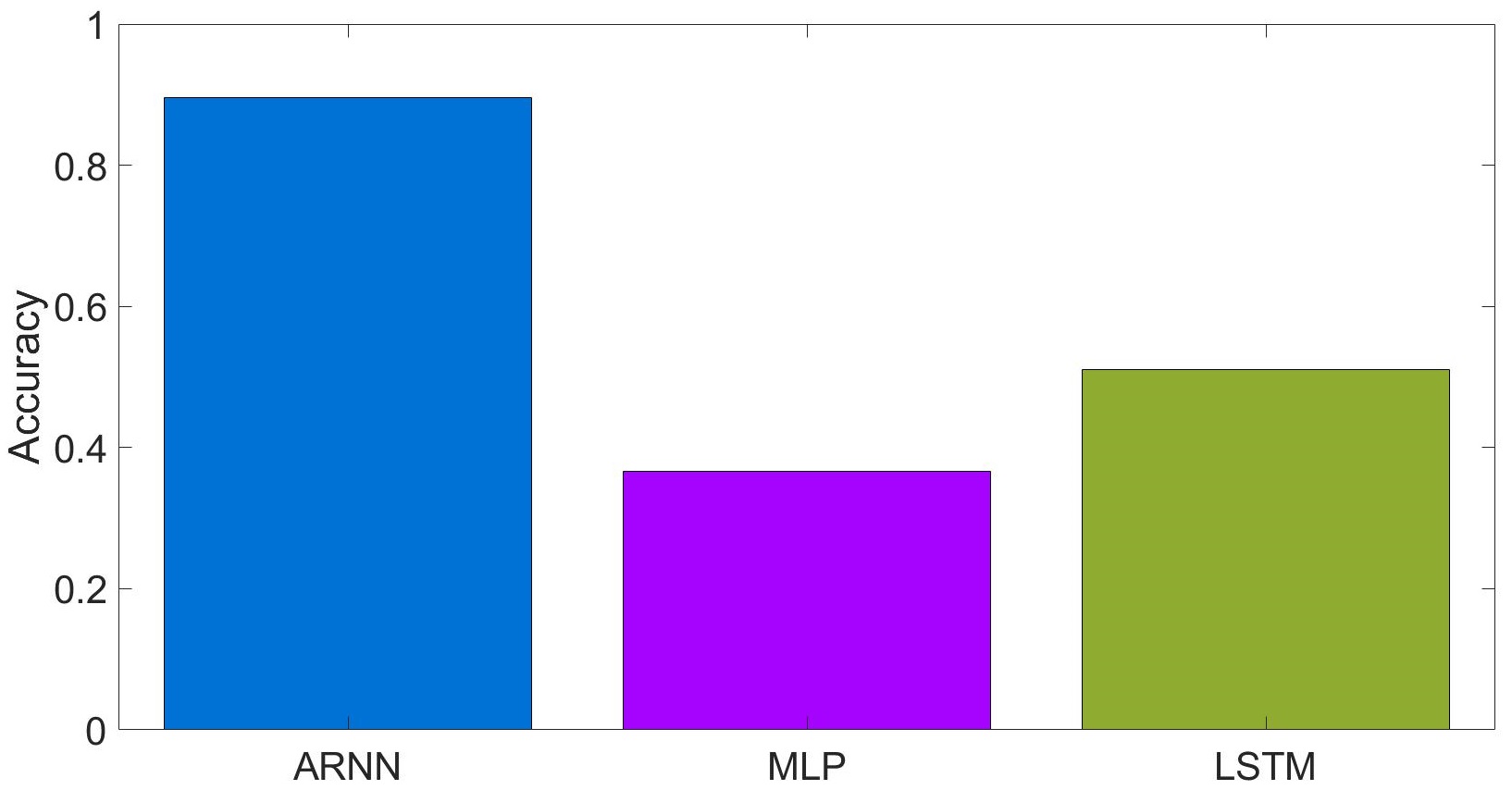}\hfill \includegraphics[width=0.475\textwidth]{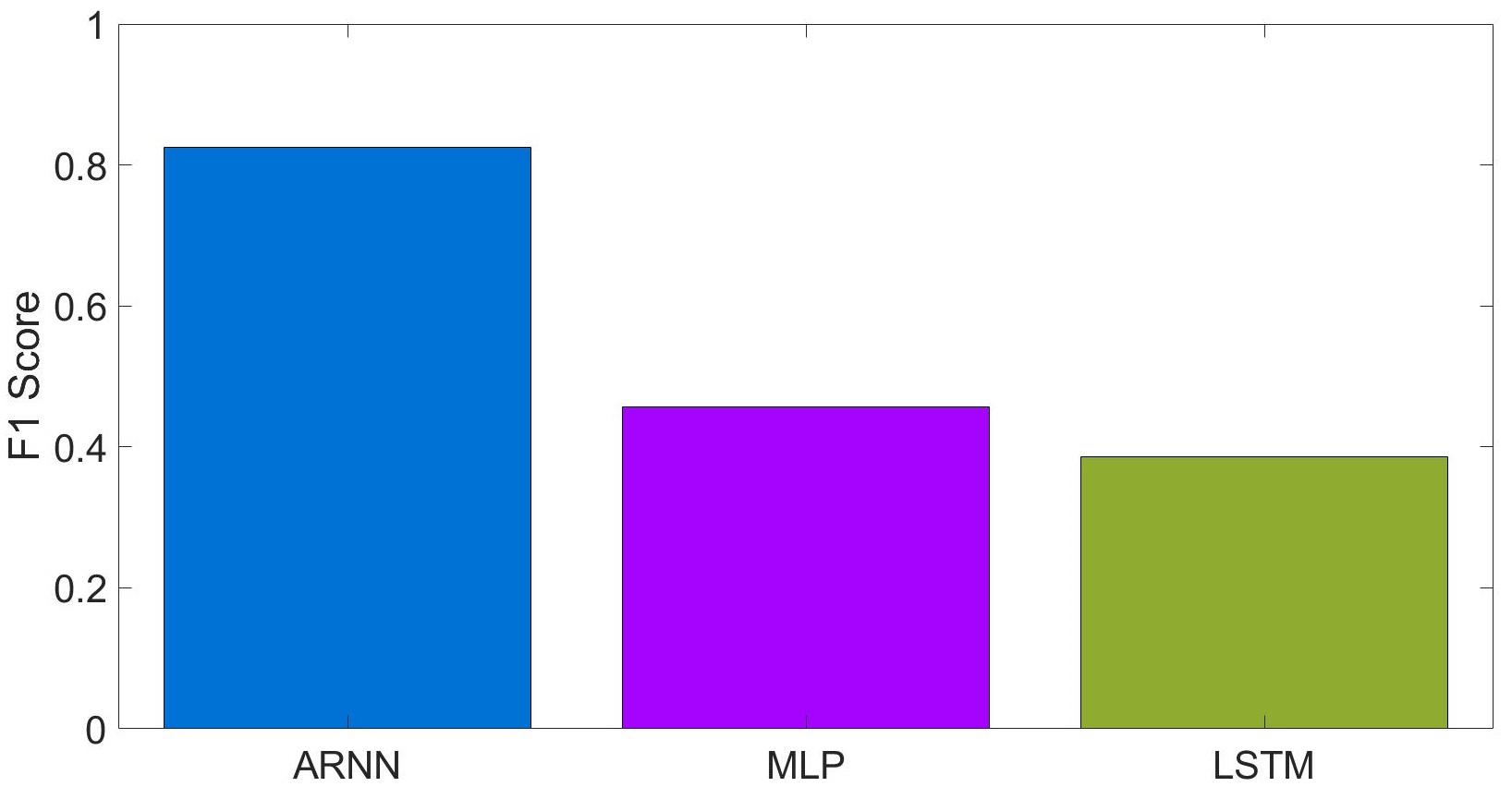}
	
	\vspace{0.5cm}
	
	\includegraphics[width=0.475\textwidth]{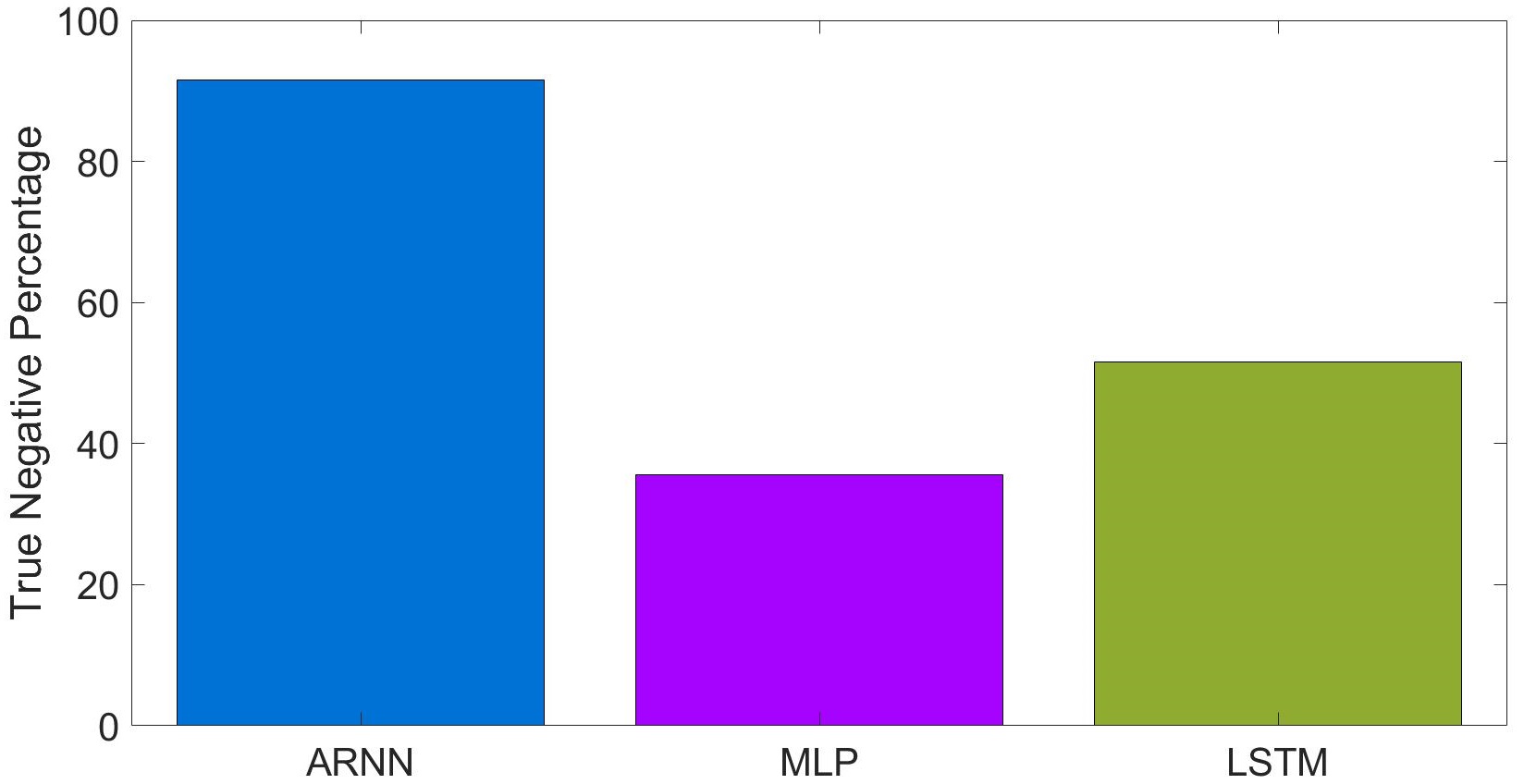}\hfill \includegraphics[width=0.475\textwidth]{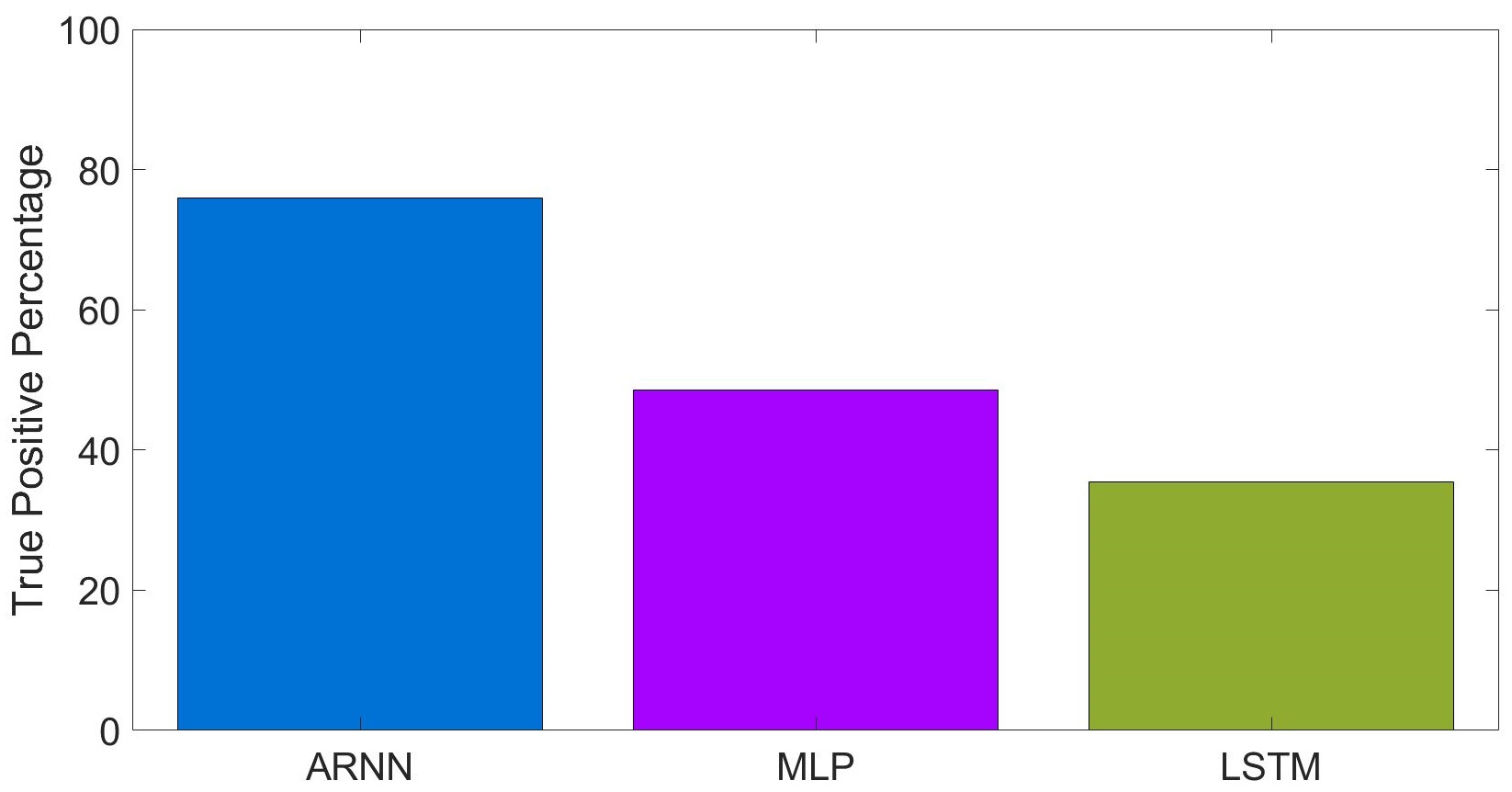}
	\caption{Performance comparison between  ARNN, MLP and LSTM for Experiment II (where each model is trained online) with respect to \textbf{(top left)} Accuracy, \textbf{(top right)} F1 Score, \textbf{(bottom left)} percentage TNR, and \textbf{(bottom right)} percentage TPR}
	\label{fig:Comparison_online}
\end{figure*}

%\subsubsection{Experiment II - Online (Incremental) Training} 
Then, in Figure~\ref{fig:Comparison_online}, the comparison of the neural network models for Experiment II (online training) with respect to the mean of each Accuracy, F1 Score, TNR and TPR is presented. The results in this figure show that  ARNN significantly outperforms both MLP and LSTM with respect to any measure by at least $27\%$. Moreover, we see that although the overall performances of both MLP and LSTM have been significantly decreased under online training compared with offline training, the performance of  ARNN is almost the same under both online and offline training. 

\subsection{Training and Execution Times} 

Finally, in Table~\ref{table:comp_time}, we present the average training and execution time. Note that these results are collected on a workstation with $32$ Gb RAM and an AMD $3.7$ GHz (Ryzen 7 3700X) processor. The second row of this table displays the average training time that has been spent for a single data sample in a single training step. Thus, during the discussion of the results on training time, we shall calculate the total training time during Experiment I and that for one training window during Experiment II. One should note that both the number of inputs and the number of outputs of  ARNN are twice those of MLP and LSTM. One should also note that the implementation of  ARNN can be optimized to achieve lower training and execution time, and both MLP and LSTM have been implemented by using Keras library in Python. 

\begin{table}[h!]
	\normalsize
	\caption{Average Training Time per Sample per Step and Average Execution Time per Sample of   ARNN, MLP and LSTM}
	\setlength{\tabcolsep}{6pt} 
	\centering
	\begin{tabular}{|c|c|c|c|}
		\hline
		          &   ARNN  & MLP    & LSTM   \\ \hline
		Training ($s$) & $40.02$ & $3.82\times10^{-4}$ & $0.01$ \\ \hline
		Execution ($ms$) & $8.4$ & $0.17$ & $0.78$ \\ \hline
	\end{tabular}
	\label{table:comp_time}
\end{table}

During Experiment I, ARNN, MLP and LSTM have been trained on 25 samples for 20 epochs, 1000 epochs and 1000 epochs, respectively. Accordingly, the total training time of these models are $40.02\times25\times20 = 20010~s$, $3.82\times10^{-4}\times25\times1000 = 9.55~s$, and $0.01\times25\times1000 = 250~s$, respectively. We see that the training time of ARNN is much higher than those of the other models. However, ARNN can be selected as identification method while the training of all models in Experiment I is performed offline and  ARNN achieves significantly higher accuracy than MLP and LSTM.

During Experiment II, all three models have been trained online on 1 minute windows (6 samples) for 3 epochs, 100 epochs and 100 epochs respectively. Accordingly, the total training time of these models for each window are $40.02\times6\times3 = 720.36~s$, $3.82\times10^{-4}\times6\times100 = 0.23~s$, and $0.01\times6\times100 = 6~s$, respectively. Although the training time results show that MLP and LSTM are suitable for training once in 1 minute, the performance of either MLP or LSTM has shown not to be acceptable for practical usage. On the other hand, ARNN with its current implementation achieves high accuracy but can be trained once in 720.36 seconds ($\approx$12 minutes) on 1 minute of data.

Furthermore, the third row of this table displays the average execution time that has been spent to make a prediction for a single sample. The results in this row show that the execution time of ARNN is one order of magnitude higher than the execution times of MLP and LSTM. 

%total training time that has been spent during the learning stage of the considered method and the average execution time that has been spent to make a prediction for single sample during the test of the considered method. In this table, the training time have been displayed for total time during Experiment I and for the time that has been spent during a single training window $l$. 

\section{Conclusions} \label{Conclusions}

In a network of  IP addresses, when n individual node is
attacked by a Botnet and becomes compromised, it can then compromise other network nodes
and turn them into attackers. Thus attacks may propagate across the system and affect other nodes
and IP addresses. There is a large prior literature regarding Botnet attacks, but most of the work has addressed attacks against a specific network node, while the collective detection of Botnet attacks has received less attention.

Thus  in this paper we have developed a ML based decision method, that identifies all the nodes of a given interconnected set of  nodes, that are compromised by a Botnet attack. The approach is based on designing an Associated Random Neural Network that incorporates two connected and recurrent Random Neural Networks (RNN), where each RNN offers a contradictory recommendation regarding whether any one of the
IP addresses or network nodes in the system are compromised. The final decision is taken by the
ARNN based on which of the two recommendations for each of the nodes appears to be stronger.
We have also developed a gradient based learning algorithm for the  ARNN, which learns based on linear-algebraic operations on the network weights. If the system is composed of $n$ IP addresses or nodes, then the
resulting learning algorithm is of time complexity $O(n^3)$ since all computations are based
on the inversion of $n\times n$ matrices.

In this paper, the  ARNN and its learning algorithm have been described and tested on real Botnet data involving some $760,000$ packets. The experimental results show that the ARNN provides very accurate predictions
of the order of $92\%$ for a $107$ node network. For comparison purposes, we have also implemented and tested two well known ML approaches for the same training and testing datasets, showing that the ARNN results provide significantly much better accuracy.

In future work, we plan to develop a generalization of the ARNN for multiple valued binary collective decision making and classification in other significant areas with datasets that contain inter-related or inter-dependent data items, such as social networks and the analysis of epidemics.

%Erol Gelenbe proposed the theoretical approach and architecture of the RNN as two interacting Associated networks linked to the nework graph, derived the learning algorithm used in this study and wrote the introduction, the theoretical development. He selected the experiments and checked the consistency of their outcomes. Mert Nakip suggesed the use of the likelihood ratio as the accuracy criterion. He proof-read the mathematical derivations and the text, implemented the algorithms, ran the experiments and produced the resulting tables.

%\conflictsofinterest{The authors declare no conflict of interest.}
\appendix

\section{Appendix: ARNN Learning Algorithm} \label{Appendix}

In this Appendix, we focus on the ARNN's learning algorithm, recalling that the ARNN is a specific ML structure based on the Random Neural Network (RNN), which has been proven to be an effective approximator in the sense of \cite{Cybenko} for continuous and bounded functions \cite{bib:GelenbeApprox1999}. It was generalized to G-Networks in the framework of queueing theory \cite{HarrisonPitel,Harrison03,Fourneau13}.
Gradient learning for the RNN was initially designed for both feedforward and recurrent (feedback) RNNs \cite{bib:Gelenbe1993}, and other RNN learning algorithms have also been proposed \cite{Rubino3,bib:TIMOTHEOU2009,gelenbe2017deepdense}.

Prior to running the learning algorithm, the  ARNN parameters are set to ``neutral'' values which express the fact that {\em initially} the ARNN does not know whether any of the network nodes are compromised. To this effect, we:
\begin{itemize}
	\item Initialize all the weights between $X_i$ and $Y_i$ to zero: $W^+_{ii}=w^+_{ii}=W^-_{ii}=w^-_{ii}=0$. 
	\item Set $W^+_{ij}=W^-_{ij}=w^+_{ij}=w^-_{ij}=0.5W$ for $i\neq j$, 
	and choose $Q_i=q_i=0.5$ to represent the perfect ignorance of the ARNN.
	\item Set the external inputs of the  ARNN to
	$\Lambda_i=\lambda_i=L(n-1),~L>0$, so that the {\em external} excitatory and inhibitory inputs are all initially set to an identical value.
	\item Keep $W$ constant in the learning procedure, and only learn $W^+_{ij},~w^+_{ij}$ for each $i\neq j$.
	\item Accordingly (\ref{Qq}) becomes: 
	\begin{eqnarray}
	&&q_i=Q_i =0.5\\
	&&= \frac{L(n-1)+0.25(n-1)W}{L(n-1) + (n-1)W + 0.25(n-1)W},\nonumber\\
	&&or~0.5= \frac{L+0.25W}{L + W +0.25W},\nonumber\\
	&&yielding~L=0.75W~.
	\end{eqnarray}
	\item Taking $W=1$ and $L=0.75$,
	all the neuron states are initialized with the values
	$Q_i=q_i=0.5,~i=1,~...~,n$.
\end{itemize}

%Each neuron in the RNN is represented by its  internal state which is a natural number, and  neurons exchange information among each other using positive (or excitatory) and negative (or inhibitory) spikes. A neuron which receives an excitatory spike will increase its internal state by $1$; a negative spike arriving to a neuron will reduce its state by $1$ if its state is positive, while the state will not change when an inhibitory spike arrives to a neuron whose state is already at zero. When a neuron's potential is  positive, we say it is ``excited'' and able to ``fire'' or send spikes at exponentially distributed intervals to other neurons. After firing each spike the neuron's internal state drops by one.

Now for any given value of the data, we use gradient descent to update the ARNN weights so as to search for a local minimum of the error ${\bf E}$ in equation (\ref{cost}). We drop the notation regarding the $l-th$ data item for simplicity, and compute ${\bf E}$'s derivative  with respect to each of the  ARNN weights:
\begin{align}
&E^{U,V}\equiv \frac{\partial{\bf E}}{\partial W^+_{U,V}}\nonumber\\
&=\sum_{i=1}^n ~[~(Q_i-K_i)Q^{U,V}_i + (q_i-1+K_i)q^{U,V}_i~], \label{grad_cost_W}\\
&E^{u,v}\equiv \frac{\partial {\bf E}}{\partial w^+_{u,v}}\nonumber\\
&= \sum_{i=1}^n~ [~(Q_i-K_i)Q^{u,v}_i + (q_i-1+K_i)q^{u,v}_i ~ ] ,\label{grad_cost_w}
\end{align}
where the derivatives of the  ARNN state values are denoted: 
\begin{eqnarray}
&&Q^{U,V}_i=\frac{\partial Q_i}{\partial W^+_{U,V}},~Q^{u,v}_i=\frac{\partial Q_i}{\partial w^+_{u,v}},\nonumber\\
&&q^{U,V}_i=\frac{\partial q_i}{\partial W^+_{U,V}},~q^{u,v}_i=\frac{\partial q_i}{\partial w^+_{u,v}}.\nonumber
\end{eqnarray}
We can then use the expressions (\ref{grad_cost_W}) and (\ref{grad_cost_w}) to update the ARNN weights iteratively for successive values of $d=1,~...~,|\textbf{TrainData}|$, with the Gradient Descent Rule with some $\eta>0$:
\begin{eqnarray}
&&W^+_{new,U,V}\leftarrow W^+_{U,V} - \eta E^{U,V}|_{(V^d,v^d)},\nonumber\\
&&w^+_{new,U,V}\leftarrow w^+_{u,v} - \eta E^{u,v}|_{(V^d,v^d)}.\label{grad2}
\end{eqnarray}

\subsection{Derivatives of the  ARNN State Probabilities}

Now consider the  ARNN with generic inputs $\Lambda=(\Lambda_1,~...~\Lambda_n)$ and $\lambda=(\lambda_1,~...~,\lambda_n)$. In order to obtain the derivatives needed for the gradient descent expression
(\ref{grad2}), we use (\ref{Qq}) to write:
\begin{eqnarray}
&&Q^{U,V}_i=\frac{Q_U}{D_V}1[i=V]+\sum_{j=1}^n \frac{W^+_{ji}}{D_i}~Q^{U,V}_j\nonumber\\ &&~~~~~~~-\sum_{j=1}^n\frac{Q_i[W-w^+_{ji}]}{D_i}~q^{U,V}_j,\label{dQ1}\\
&&q^{U,V}_i= \sum_{j=1}^n \frac{w^+_{ji}}{d_i}~q^{U,V}_j-\sum_{j=1}^n\frac{q_i[W-W^+_{ji}]}{d_i}~Q^{U,V}_j\nonumber\\
&&~~~~~~~+\frac{q_U}{d_V}1[i=V],\label{dq1}
\end{eqnarray}
where $D_i$  and $d_i$ are the denominators of $Q_i$ and $q_i$ respectively, in (\ref{Qq}):
\begin{eqnarray}
&&D_i=\Lambda_i+\sum_{j=1,j\neq i}^nW+\sum_{j=1,j\neq i}^n[W-w^+_{ji}]~.q_j,\\
&&d_i=\lambda_i+ \sum_{j=1,j\neq i}^nW+\sum_{j=1,j\neq i}^n[W-W^+_{ji}]~.Q_j.\nonumber
\end{eqnarray}
Define the vectors $Q=(Q_1,~...~,Q_n)$ and $q=(q_1,~...~,q_n)$ and the corresponding vectors of derivatives
$Q^{U,V}=(Q^{U,V}_1,~...~,Q^{U,V}_n)$ and $q^{U,V}=(q^{U,V}_1,~...~,q^{U,V}_n)$.
Similarly we define the $n\times n$ matrices:
\begin{eqnarray}
&&B^+=\{ \frac{W^+_{ij}}{D_j} \},~C=\{ \frac{Q_j[W-w^+_{ij}]}{D_j} \},\\
&&F^+=\{ \frac{w^+_{ij}}{d_j} \},~G=\{\frac{q_j[W-W^+_{ij}]}{d_j} \}.\nonumber
\end{eqnarray}
We use the vector $\delta_{V}$ whose elements are zero everywhere, except in position $V$ where the value is $1$, and write (\ref{dQ1}) and (\ref{dq1}) in vector form:
\begin{eqnarray}
Q^{U,V}&=&B^+Q^{U,V}-Cq^{U,V}+\delta_{V}.\frac{Q_U}{D_V},\nonumber\\
q^{U,V}&=&F^+ q^{U,V}-GQ^{U,V}+\frac{q_U}{d_V}\delta_V,\nonumber\\
&=&[-GQ^{U,V}+\frac{q_U}{d_V}\delta_V][I-F^+]^{-1}, \label{dqUV}
\end{eqnarray}
which yields:
\begin{equation}
Q^{U,V}=B^+Q^{U,V}+[CGQ^{U,V}-\frac{q_U}{d_V}C\delta_V][1-F^+]^{-1}+\delta_{V}.\frac{Q_U}{D_V},\nonumber
\end{equation}
and hence:
\begin{eqnarray}\label{dQUV}
&&Q^{U,V}=\{-\frac{q_U}{d_V}C\delta_V[I-F^+]^{-1}+\frac{Q_U}{D_V}\delta_V\}{\bf .}\nonumber\\
&&~~~~~~~~{\bf .}\{I-B^+-CG[I-F^+]^{-1}\}^{-1}\qquad
\end{eqnarray}
Also define the matrices:
\begin{eqnarray}
&&B_*^+=\{\frac{w^+_{ij}}{d_j} \},~C_*=\{\frac{q_j[W-W^+_{ij}]}{d_j} \},\\
&&F_*^+=\{ \frac{W^+_{ij}}{D_j} \},~G_*=\{\frac{Q_j[W-w^+_{ij}]}{D_j}\}.\nonumber
\end{eqnarray}
Since $Q^{U,V}$ and $q^{u,v}$ are symmetric with respect to each other, as are $Q^{u,v}$ and $q^{U,V}$, we also obtain:
\begin{eqnarray}
&&q^{u,v}=\{-\frac{Q_u}{D_v}C_*\delta_v[I-F_*^+]^{-1}+\frac{q_u}{d_v}\delta_v\}{\bf .}\nonumber\\
&&~~~~~~{\bf .}\{I-B_*^+-C_*G_*[I-F_*^+]^{-1}\}^{-1},\label{dquv}\end{eqnarray}
and
\begin{equation}
Q^{u,v}=\{-G_*q^{u,v}+\frac{Q_u}{D_v}\delta_v\}[I-F_*^+]^{-1}~.\label{dQuv}
\end{equation}
This completes the computation of all the needed derivatives of the  ARNN state probability vectors $Q$ and $q$.

\bibliographystyle{elsarticle-num}
\bibliography{references1,mybibliography,references,references12}

\begin{thebibliography}{10}
\expandafter\ifx\csname url\endcsname\relax
  \def\url#1{\texttt{#1}}\fi
\expandafter\ifx\csname urlprefix\endcsname\relax\def\urlprefix{URL }\fi
\expandafter\ifx\csname href\endcsname\relax
  \def\href#1#2{#2} \def\path#1{#1}\fi

\bibitem{Davis}
W.~R. Schwartz, H.~Guo, J.~Choi, L.~S. Davis, Face identification using large
  feature sets, IEEE Transactions on Image Processing 21~(4) (2012) 2245--2255.
\newblock \href {https://doi.org/10.1109/TIP.2011.2176951}
  {\path{doi:10.1109/TIP.2011.2176951}}.

\bibitem{Binary}
A.~Ortner,
  \href{https://github.com/alexortner/teaching/tree/master/binary_classification}{Top
  10 binary classification algorithms [a beginner’s guide]} (May 2020).
\newline\urlprefix\url{https://github.com/alexortner/teaching/tree/master/binary_classification}

\bibitem{Botnet1}
T.~A. Tuan, H.~V. Long, L.~H. Son, R.~Kumar, I.~Priyadarshini, N.~T.~K. Son,
  \href{https://doi.org/10.1007/s12065-019-00310-w}{Performance evaluation of
  botnet ddos attack detection using machine learning}, {Evolutionary
  Intelligence} (October 2019).
\newline\urlprefix\url{https://doi.org/10.1007/s12065-019-00310-w}

\bibitem{Filus21}
K.~Filus, J.~Doma{\'n}ska, E.~Gelenbe, Random neural network for lightweight
  attack detection in the {IoT}, in: Symposium on Modelling, Analysis, and
  Simulation of Computer and Telecommunication Systems, Springer, Cham, 2021,
  pp. 79--91.

\bibitem{Alshamkhany}
M.~Alshamkhany, W.~Alshamkhany, M.~Mansour, M.~Khan, S.~Dhou, F.~Aloul, Botnet
  attack detection using machine learning, in: 14th International Conference on
  Innovations in Information Technology (IIT), 2020, pp. 203--208.
\newblock \href {https://doi.org/10.1109/IIT50501.2020.9299061}
  {\path{doi:10.1109/IIT50501.2020.9299061}}.

\bibitem{Access22}
E.~Gelenbe, M.~Nakıp, Traffic based sequential learning during {Botnet}
  attacks to identify compromised iot devices, IEEE Access 10 (2022)
  126536--126549.
\newblock \href {https://doi.org/10.1109/ACCESS.2022.3226700}
  {\path{doi:10.1109/ACCESS.2022.3226700}}.

\bibitem{Collective1}
M.~Bilgic, G.~M. Namata, L.~Getoor, Combining collective classification and
  link prediction, in: Seventh IEEE International Conference on Data Mining
  Workshops (ICDMW 2007), 2007, pp. 381--386.
\newblock \href {https://doi.org/10.1109/ICDMW.2007.35}
  {\path{doi:10.1109/ICDMW.2007.35}}.

\bibitem{Collective3}
P.~Sen, G.~Namata, M.~Bilgic, L.~Getoor,
  \href{https://doi.org/10.1007/978-0-387-30164-8_140}{{Collective
  Classification}}, in: C.~Sammut, G.~I. Webb (Eds.), Encyclopedia of Machine
  Learning, Springer US, Boston, MA, 2010, pp. 189--193.
\newblock \href {https://doi.org/10.1007/978-0-387-30164-8_140}
  {\path{doi:10.1007/978-0-387-30164-8_140}}.
\newline\urlprefix\url{https://doi.org/10.1007/978-0-387-30164-8_140}

\bibitem{Collective2}
P.~Sen, G.~Namata, M.~Bilgic, L.~Getoor, B.~Galligher, T.~Eliassi-Rad,
  \href{https://ojs.aaai.org/index.php/aimagazine/article/view/2157}{Collective
  classification in network data}, AI Magazine 29~(3) (2008) 93.
\newblock \href {https://doi.org/10.1609/aimag.v29i3.2157}
  {\path{doi:10.1609/aimag.v29i3.2157}}.
\newline\urlprefix\url{https://ojs.aaai.org/index.php/aimagazine/article/view/2157}

\bibitem{Collective4}
B.~London, L.~Getoor, Collective classification of network data, Data
  Classification: Algorithms and Applications 29 (2014) 399–416.

\bibitem{Collective5}
T.~Kajdanowicz, P.~Kazienko, Collective classification, in: {Encyclopedia of
  Social Network Analysis and Mining}, 2018, p. 253–265.
\newblock \href {https://doi.org/10.1007/978-1-4939-7131-2_45. ISBN
  978-1-4939-7130-5} {\path{doi:10.1007/978-1-4939-7131-2_45. ISBN
  978-1-4939-7130-5}}.

\bibitem{CollectiveSurvey1}
M.~Bailey, E.~Cooke, F.~Jahanian, Y.~Xu, M.~Karir, A survey of botnet
  technology and defenses.

\bibitem{CollectiveSurvey2}
S.~García, A.~Zunino, M.~Campo, Survey on network-based botnet detection
  methods, Secur. Commun. Netw.~(7) (2014) 878–903.

\bibitem{CollectiveSurvey3}
H.~Owen, J.~Zarrin, S.~M. Pour, \href{https://www.mdpi.com/2624-800X/2/1/6}{A
  survey on botnets, issues, threats, methods, detection and prevention},
  Journal of Cybersecurity and Privacy 2~(1) (2022) 74--88.
\newblock \href {https://doi.org/10.3390/jcp2010006}
  {\path{doi:10.3390/jcp2010006}}.
\newline\urlprefix\url{https://www.mdpi.com/2624-800X/2/1/6}

\bibitem{CollectiveDetect1}
G.~Gu, J.~Zhang, W.~Lee, Botsniffer: Detecting botnet command and control
  channels in network traffic, in: Proceedings of the 15th Annual Network and
  Distributed System Security Symposium (NDSS’08).

\bibitem{CollectiveDetect2}
H.~Joshi, M.~Bennison, R.~Dutta, Collaborative botnet detection with partial
  communication graph information, in: Proceedings of the 2017 IEEE 38th
  Sarnoff Symposium.

\bibitem{CollectiveDetect3}
Z.~Yang, B.~Wang, \href{https://www.mdpi.com/2078-2489/10/5/160}{P2p botnet
  detection based on nodes correlation by the mahalanobis distance},
  Information 10~(5) (2019).
\newblock \href {https://doi.org/10.3390/info10050160}
  {\path{doi:10.3390/info10050160}}.
\newline\urlprefix\url{https://www.mdpi.com/2078-2489/10/5/160}

\bibitem{RNN1}
E.~Gelenbe, Random neural networks with negative and positive signals and
  product form solution, Neural Computation 1~(4) (1989) 502--510.

\bibitem{Video}
C.~E. Cramer, E.~Gelenbe, Video quality and traffic qos in learning-based
  subsampled and receiver-interpolated video sequences, IEEE Journal on
  Selected Areas in Communications 18~(2) (2000) 150--167.

\bibitem{Aiello}
G.~Aiello, S.~Gaglio, G.~Lo-Re, P.~Storniolo, A.~Urso,
  \href{https://doi.org/10.1007/1-4020-3432-6_19}{The random neural network
  model for the on-line multicast problem}, in: B.~Apolloni, M.~Marinaro, T.~R.
  (Eds.), Biological and Artificial Intelligence Environments, Springer,
  Dordrecht, 2005.
\newline\urlprefix\url{https://doi.org/10.1007/1-4020-3432-6_19}

\bibitem{Khaled0}
E.~Gelenbe, K.~F. Hussain, Learning in the multiple class random neural
  network, IEEE Transactions on Neural Networks 13~(6) (2002) 1257--1267.

\bibitem{Kaptan1}
K.~F. Hussain, V.~Kaptan, Modeling and simulation with augmented reality, Int.
  J. Oper. Res. 38~(2) (2004) 89--103.

\bibitem{Khaled1}
K.~F. Hussain, G.~S. Moussa, On road vehicle classification based on random
  neural network and bag of visual words, Probability in the Engineering and
  Informational Sciences 30 (2016) 403--412.
\newblock \href {https://doi.org/doi:10.1017/S0269964816000073}
  {\path{doi:doi:10.1017/S0269964816000073}}.

\bibitem{Khaled2}
K.~F. Hussain, E.~Radwan, G.~S. Moussa, Augmented reality experiment: Drivers'
  behavior at an unsignalized intersection, IEEE Transactions on Intelligent
  Transportation Systems 14~(2) (2013) 608--617.

\bibitem{BuildingEnergy2}
J.~Ahmad, A.~Tahir, H.~Larijani, F.~Ahmed, S.~A. Shah, A.~J. Hall, W.~J.
  Buchanan, \href{https://doi.org/10.3233/JIFS-191458}{Energy demand
  forecasting of buildings using random neural networks}, J. Intell. Fuzzy
  Syst. 38~(4) (2020) 4753--4765.
\newline\urlprefix\url{https://doi.org/10.3233/JIFS-191458}

\bibitem{VideoScheduling}
T.~Ghalut, H.~Larijani,
  \href{https://doi.org/10.5383/JUSPN.09.02.003}{Content-aware and {QOE}
  optimization of video stream scheduling over {LTE} networks using genetic
  algorithms and random neural networks}, J. Ubiquitous Syst. Pervasive
  Networks 9~(2) (2018) 21--33.
\newline\urlprefix\url{https://doi.org/10.5383/JUSPN.09.02.003}

\bibitem{Intrusion}
A.~Qureshi, H.~Larijani, J.~Ahmad, N.~Mtetwa,
  \href{https://doi.org/10.1109/CEEC.2018.8674228}{A novel random neural
  network based approach for intrusion detection systems}, in: 2018 10th
  Computer Science and Electronic Engineering Conference, {CEEC} 2018,
  University of Essex, Colchester, UK, September 19-21, 2018, {IEEE}, 2018, pp.
  50--55.
\newline\urlprefix\url{https://doi.org/10.1109/CEEC.2018.8674228}

\bibitem{AdaptiveModulation}
A.~Adeel, H.~Larijani, A.~Ahmadinia,
  \href{https://doi.org/10.1016/j.compeleceng.2016.11.005}{Random neural
  network based cognitive engines for adaptive modulation and coding in {LTE}
  downlink systems}, Comput. Electr. Eng. 57 (2017) 336--350.
\newline\urlprefix\url{https://doi.org/10.1016/j.compeleceng.2016.11.005}

\bibitem{HVAC1}
A.~Javed, H.~Larijani, A.~Ahmadinia, R.~Emmanuel, M.~Mannion, D.~Gibson,
  \href{https://doi.org/10.1109/JIOT.2016.2627403}{Design and implementation of
  a cloud enabled random neural network-based decentralized smart controller
  with intelligent sensor nodes for {HVAC}}, {IEEE} Internet Things J. 4~(2)
  (2017) 393--403.
\newline\urlprefix\url{https://doi.org/10.1109/JIOT.2016.2627403}

\bibitem{HVAC2}
A.~Javed, H.~Larijani, A.~Ahmadinia, D.~Gibson,
  \href{https://doi.org/10.1109/TII.2016.2597746}{Smart random neural network
  controller for {HVAC} using cloud computing technology}, {IEEE} Trans. Ind.
  Informatics 13~(1) (2017) 351--360.
\newline\urlprefix\url{https://doi.org/10.1109/TII.2016.2597746}

\bibitem{BuildingEnergy1}
J.~Ahmad, H.~Larijani, R.~Emmanuel, M.~Mannion, A.~Javed, M.~Phillipson,
  \href{https://doi.org/10.1109/SYSCON.2017.7934803}{Energy demand prediction
  through novel random neural network predictor for large non-domestic
  buildings}, in: 2017 Annual {IEEE} International Systems Conference, SysCon
  2017, Montreal, QC, Canada, April 24-27, 2017, {IEEE}, 2017, pp. 1--6.
\newline\urlprefix\url{https://doi.org/10.1109/SYSCON.2017.7934803}

\bibitem{VoiceQuality}
K.~Radhakrishnan, H.~Larijani,
  \href{https://doi.org/10.1016/j.peva.2011.01.001}{Evaluating perceived voice
  quality on packet networks using different random neural network
  architectures}, Perform. Evaluation 68~(4) (2011) 347--360.
\newline\urlprefix\url{https://doi.org/10.1016/j.peva.2011.01.001}

\bibitem{VideoQuality}
T.~Ghalut, H.~Larijani,
  \href{https://doi.org/10.1109/CSNDSP.2014.6923884}{Non-intrusive method for
  video quality prediction over {LTE} using random neural networks {(RNN)}},
  in: 9th International Symposium on Communication Systems, Networks {\&}
  Digital Signal Processing, {CSNDSP} 2014, Manchester, UK, July 23-25, 2014,
  {IEEE}, 2014, pp. 519--524.
\newline\urlprefix\url{https://doi.org/10.1109/CSNDSP.2014.6923884}

\bibitem{Virus}
E.~Gelenbe, Dealing with software viruses: a biological paradigm, Information
  Security Technical Report 12~(4) (2007) 242--250.

\bibitem{Rubino1}
G.~Rubino, P.~Tirilly, M.~Varela,
  \href{https://doi.org/10.1007/11840817\_32}{Evaluating users' satisfaction in
  packet networks using random neural networks}, in: S.~D. Kollias,
  A.~Stafylopatis, W.~Duch, E.~Oja (Eds.), Artificial Neural Networks - {ICANN}
  2006, 16th International Conference, Athens, Greece, September 10-14, 2006.
  Proceedings, Part {I}, Vol. 4131 of Lecture Notes in Computer Science,
  Springer, 2006, pp. 303--312.
\newline\urlprefix\url{https://doi.org/10.1007/11840817\_32}

\bibitem{Rubino2}
M.~Martínez, A.~Mor\`{o}n, F.~Robledo, P.~Rodríguez-Bocca, H.~Cancela,
  G.~Rubino, A grasp algorithm using rnn for solving dynamics in a p2p live
  video streaming network, in: 2008 Eighth International Conference on Hybrid
  Intelligent Systems, Barcelona, 2008, pp. 447--452.
\newblock \href {https://doi.org/10.1109/HIS.2008.23}
  {\path{doi:10.1109/HIS.2008.23}}.

\bibitem{Rubino3}
S.~Basterrech, S.~Mohamed, G.~Rubino, M.~A. Soliman,
  \href{https://doi.org/10.1093/comjnl/bxp101}{Levenberg-marquardt training
  algorithms for random neural networks}, Comput. J. 54~(1) (2011) 125--135.
\newline\urlprefix\url{https://doi.org/10.1093/comjnl/bxp101}

\bibitem{Yin1}
I.~Grenet, Y.~Yin, J.-P. Comet, G-networks to predict the outcome of sensing of
  toxicity, Sensors~(10) (2018) 3483.

\bibitem{Yin4}
Y.~Yin, Random neural network methods and deep learning, {Probability in the
  Engineering and Informational Sciences} 34~(1) (2021) 6--35.

\bibitem{KitsuneKaggle}
\href{https://www.kaggle.com/ymirsky/network-attack-dataset-kitsune}{{Kitsune
  Network Attack Dataset}} (August 2020).
\newline\urlprefix\url{https://www.kaggle.com/ymirsky/network-attack-dataset-kitsune}

\bibitem{bib:ciscopriv}
{Cisco Cybersecurity Series 2019. Consumer Privacy Survey}, [online],
  available:
  \url{https://www.cisco.com/c/dam/en_us/about/annual-report/cisco-annual-report-2019.pdf}
  [Accessed: 2020-08-05] (Cisco 2019).

\bibitem{bib:cisco}
Cisco,
  \href{https://www.cisco.com/c/dam/en_us/about/annual-report/2018-annual-report-full.pdf}{Cisco
  2018 annual cybersecurity report} (2018).
\newline\urlprefix\url{https://www.cisco.com/c/dam/en_us/about/annual-report/2018-annual-report-full.pdf}

\bibitem{HTTP}
O.~Drexler,
  \href{https://www.clickguard.com/blog/recent-botnet-attacks-2022/}{{The Most
  Recent Botnet Attacks: The 2022 Edition}}.
\newline\urlprefix\url{https://www.clickguard.com/blog/recent-botnet-attacks-2022/}

\bibitem{Survey}
S.~N. Thanh~Vu, M.~Stege, P.~I. El-Habr, J.~Bang, N.~Dragoni,
  \href{https://www.mdpi.com/1999-5903/13/8/198}{A survey on botnets:
  Incentives, evolution, detection and current trends}, Future Internet 13~(8)
  (2021).
\newblock \href {https://doi.org/10.3390/fi13080198}
  {\path{doi:10.3390/fi13080198}}.
\newline\urlprefix\url{https://www.mdpi.com/1999-5903/13/8/198}

\bibitem{Botnet25}
X.~Dong, J.~Hu, Y.~Cui, Overview of botnet detection based on machine learning,
  in: 2018 3rd International Conference on Mechanical, Control and Computer
  Engineering (ICMCCE), 2018, pp. 476--479.
\newblock \href {https://doi.org/10.1109/ICMCCE.2018.00106}
  {\path{doi:10.1109/ICMCCE.2018.00106}}.

\bibitem{Botnet2}
S.~N. Thanh~Vu, M.~Stege, P.~I. El-Habr, J.~Bang, N.~Dragoni,
  \href{https://www.mdpi.com/1999-5903/13/8/198}{A survey on botnets:
  Incentives, evolution, detection and current trends}, {Future Internet}
  13~(8) (2021).
\newblock \href {https://doi.org/10.3390/fi13080198}
  {\path{doi:10.3390/fi13080198}}.
\newline\urlprefix\url{https://www.mdpi.com/1999-5903/13/8/198}

\bibitem{Mirai}
CLOUDFLARE,
  \href{https://www.cloudflare.com/learning/ddos/glossary/mirai-botnet/}{What
  is the mirai botnet} (December 2022).
\newline\urlprefix\url{https://www.cloudflare.com/learning/ddos/glossary/mirai-botnet/}

\bibitem{Meris1}
V.~Ganti, O.~Yoachimik, \href{https://blog.cloudflare.com/meris-botnet/}{A
  brief history of the meris botnet} (September 2021).
\newline\urlprefix\url{https://blog.cloudflare.com/meris-botnet/}

\bibitem{Meris2}
\href{https://www.cshub.com/attacks/news/google-blocks-largest-ever-web-ddos-attack}{Google
  blocks ``largest ever'' web ddos attack}.
\newline\urlprefix\url{https://www.cshub.com/attacks/news/google-blocks-largest-ever-web-ddos-attack}

\bibitem{bib:Jeatrakul2009}
P.~{Jeatrakul}, K.~W. {Wong}, Comparing the performance of different neural
  networks for binary classification problems, in: 2009 Eighth International
  Symposium on Natural Language Processing, 2009, pp. 111--115.

\bibitem{Botnet12}
W.~Zhang, Y.-J. Wang, X.-L. Wang, A survey of defense against p2p botnets, in:
  2014 IEEE 12th International Conference on Dependable, Autonomic and Secure
  Computing, 2014, pp. 97--102.
\newblock \href {https://doi.org/10.1109/DASC.2014.26}
  {\path{doi:10.1109/DASC.2014.26}}.

\bibitem{Botnet8}
H.~Dhayal, J.~Kumar, Botnet and p2p botnet detection strategies: A review, in:
  2018 International Conference on Communication and Signal Processing (ICCSP),
  2018, pp. 1077--1082.
\newblock \href {https://doi.org/10.1109/ICCSP.2018.8524529}
  {\path{doi:10.1109/ICCSP.2018.8524529}}.

\bibitem{bib:yin2017deep}
C.~Yin, Y.~Zhu, J.~Fei, X.~He, A deep learning approach for intrusion detection
  using recurrent neural networks, IEEE Access 5 (2017) 21954--21961.

\bibitem{bib:cortes2019hybrid}
F.~M. Cort{\'e}s, N.~G. G{\'o}mez, A hybrid alarm management strategy in
  signature-based intrusion detection systems, in: 2019 IEEE Colombian
  Conference on Communications and Computing (COLCOM), 2019, pp. 1--6.

\bibitem{bib:li2019designing}
W.~Li, S.~Tug, W.~Meng, Y.~Wang, Designing collaborative blockchained
  signature-based intrusion detection in iot environments, Future Generation
  Computer Systems 96 (2019) 481--489.

\bibitem{bib:medbiotguerra2020medbiot}
A.~Guerra-Manzanares, J.~Medina-Galindo, H.~Bahsi, S.~N{\~o}mm, {MedBIoT}:
  Generation of an {IoT} botnet dataset in a medium-sized {IoT} network., in:
  6th International Conference on Information Systems Security and Privacy,
  2020, pp. 207--218.

\bibitem{Botnet3}
M.~H.~B. Kamilin, S.~Yamaguchi, White-hat worm launcher based on deep learning
  in botnet defense system, in: 2020 IEEE International Conference on Consumer
  Electronics - Asia (ICCE-Asia), 2020, pp. 1--2.
\newblock \href {https://doi.org/10.1109/ICCE-Asia49877.2020.9277358}
  {\path{doi:10.1109/ICCE-Asia49877.2020.9277358}}.

\bibitem{Botnet7}
S.~Yamaguchi, A basic command and control strategy in botnet defense system,
  in: 2021 IEEE International Conference on Consumer Electronics (ICCE), 2021,
  pp. 1--5.
\newblock \href {https://doi.org/10.1109/ICCE50685.2021.9427667}
  {\path{doi:10.1109/ICCE50685.2021.9427667}}.

\bibitem{Botnet4}
S.-C. Chen, Y.-R. Chen, W.-G. Tzeng, Effective botnet detection through neural
  networks on convolutional features, in: 2018 17th IEEE International
  Conference On Trust, Security And Privacy In Computing And Communications/
  12th IEEE International Conference On Big Data Science And Engineering
  (TrustCom/BigDataSE), 2018, pp. 372--378.
\newblock \href {https://doi.org/10.1109/TrustCom/BigDataSE.2018.00062}
  {\path{doi:10.1109/TrustCom/BigDataSE.2018.00062}}.

\bibitem{Botnet5}
B.~Alothman, P.~Rattadilok, Towards using transfer learning for botnet
  detection, in: 2017 12th International Conference for Internet Technology and
  Secured Transactions (ICITST), 2017, pp. 281--282.
\newblock \href {https://doi.org/10.23919/ICITST.2017.8356400}
  {\path{doi:10.23919/ICITST.2017.8356400}}.

\bibitem{Botnet6}
G.~Vormayr, T.~Zseby, J.~Fabini, Botnet communication patterns, IEEE
  Communications Surveys Tutorials 19~(4) (2017) 2768--2796.
\newblock \href {https://doi.org/10.1109/COMST.2017.2749442}
  {\path{doi:10.1109/COMST.2017.2749442}}.

\bibitem{Botnet19}
J.~Pijpker, H.~Vranken, The role of internet service providers in botnet
  mitigation, in: 2016 European Intelligence and Security Informatics
  Conference (EISIC), 2016, pp. 24--31.
\newblock \href {https://doi.org/10.1109/EISIC.2016.013}
  {\path{doi:10.1109/EISIC.2016.013}}.

\bibitem{Botnet10}
A.~Woodiss-Field, M.~N. Johnstone, Assessing the suitability of traditional
  botnet detection against contemporary threats, in: 2020 Workshop on Emerging
  Technologies for Security in IoT (ETSecIoT), 2020, pp. 18--21.
\newblock \href {https://doi.org/10.1109/ETSecIoT50046.2020.00008}
  {\path{doi:10.1109/ETSecIoT50046.2020.00008}}.

\bibitem{Botnet17}
S.~Y. Yerima, M.~K. Alzaylaee, Mobile botnet detection: A deep learning
  approach using convolutional neural networks, in: 2020 International
  Conference on Cyber Situational Awareness, Data Analytics and Assessment
  (CyberSA), 2020, pp. 1--8.
\newblock \href {https://doi.org/10.1109/CyberSA49311.2020.9139664}
  {\path{doi:10.1109/CyberSA49311.2020.9139664}}.

\bibitem{Botnet22}
M.~Thangapandiyan, P.~M.~R. Anand, An efficient botnet detection system for p2p
  botnet, in: 2016 International Conference on Wireless Communications, Signal
  Processing and Networking (WiSPNET), 2016, pp. 1217--1221.
\newblock \href {https://doi.org/10.1109/WiSPNET.2016.7566330}
  {\path{doi:10.1109/WiSPNET.2016.7566330}}.

\bibitem{Botnet21}
M.~T. Garip, J.~Lin, P.~Reiher, M.~Gerla, Shieldnet: An adaptive detection
  mechanism against vehicular botnets in vanets, in: 2019 IEEE Vehicular
  Networking Conference (VNC), 2019, pp. 1--7.
\newblock \href {https://doi.org/10.1109/VNC48660.2019.9062790}
  {\path{doi:10.1109/VNC48660.2019.9062790}}.

\bibitem{Botnet24}
Y.~Aleksieva, H.~Valchanov, V.~Aleksieva, An approach for host based {B}otnet
  detection system, in: 2019 16th Conference on Electrical Machines, Drives and
  Power Systems (ELMA), 2019, pp. 1--4.
\newblock \href {https://doi.org/10.1109/ELMA.2019.8771644}
  {\path{doi:10.1109/ELMA.2019.8771644}}.

\bibitem{Botnet26}
F.~Hussain, S.~G. Abbas, U.~U. Fayyaz, G.~A. Shah, A.~Toqeer, A.~Ali, Towards a
  universal features set for iot botnet attacks detection, in: 2020 IEEE 23rd
  International Multitopic Conference (INMIC), 2020, pp. 1--6.
\newblock \href {https://doi.org/10.1109/INMIC50486.2020.9318106}
  {\path{doi:10.1109/INMIC50486.2020.9318106}}.

\bibitem{Botnet23}
Y.~Chen, B.~Pang, G.~Shao, G.~Wen, X.~Chen, Dga-based botnet detection toward
  imbalanced multiclass learning, Tsinghua Science and Technology 26~(4) (2021)
  387--402.
\newblock \href {https://doi.org/10.26599/TST.2020.9010021}
  {\path{doi:10.26599/TST.2020.9010021}}.

\bibitem{Spilios}
S.~Evmorfos, G.~Vlachodimitropoulos, N.~Bakalos, E.~Gelenbe, Neural network
  architectures for the detection of syn flood attacks in iot systems, in:
  Proceedings of the 13th ACM International Conference on PErvasive
  Technologies Related to Assistive Environments, no.~69, ACM;
  https://doi.org/10.1145/3389189.3398000, 2020, pp. 1--4.

\bibitem{Gelenbe2020}
E.~Gelenbe, J.~Domanska, P.~Fr\"{o}hlich, M.~Nowak, S.~Nowak, Self-aware
  networks that optimize security, qos and energy, Proceedings of the IEEE
  108~(7) (2020) 1150--1167.
\newblock \href {https://doi.org/10.1109/JPROC.2020.2992559}
  {\path{doi:10.1109/JPROC.2020.2992559}}.

\bibitem{gelenbe2017deepdense}
E.~Gelenbe, Y.~Yin, Deep learning with dense random neural networks, in:
  International Conference on Man--Machine Interactions, Springer, Cham, 2017,
  pp. 3--18.

\bibitem{Brun}
O.~Brun, Y.~Yin, E.~Gelenbe,
  \href{https://doi.org/10.1016/j.procs.2018.07.183}{Deep learning with dense
  random neural network for detecting attacks against iot-connected home
  environments}, Procedia Computer Science 134 (2018) 458--463.
\newline\urlprefix\url{https://doi.org/10.1016/j.procs.2018.07.183}

\bibitem{Nakip}
M.~Nakip, E.~Gelenbe, Mirai botnet attack detection with auto-associative dense
  random neural network, in: 2021 IEEE Global Communications Conference
  (GLOBECOM), 2021, pp. 01--06.
\newblock \href {https://doi.org/10.1109/GLOBECOM46510.2021.9685306}
  {\path{doi:10.1109/GLOBECOM46510.2021.9685306}}.

\bibitem{Nakip_incremental}
M.~Nakip, E.~Gelenbe, Botnet attack detection with incremental online learning,
  in: Security in Computer and Information Sciences: Second International
  Symposium, EuroCybersec 2021, Nice, France, October 25--26, 2021, Revised
  Selected Papers, Springer, 2022, pp. 51--60.

\bibitem{G-Nets-Attack}
E.~Gelenbe, M.~Nakip, G-networks can detect different types of cyberattacks.

\bibitem{bib:Gelenbe1993}
E.~Gelenbe, Learning in the recurrent random neural network, Neural Computation
  5 (1993) 154--164.

\bibitem{mirsky2018kitsune}
Y.~Mirsky, T.~Doitshman, Y.~Elovici, A.~Shabtai, Kitsune: An ensemble of
  autoencoders for online network intrusion detection, in: The Network and
  Distributed System Security Symposium (NDSS) 2018, 2018.

\bibitem{Soldatos2}
J.~Soldatos, S.~A. Kyriazakos, P.~Ziafati, A.~D. Mihovska, Securing iot
  applications with smart objects: Framework and a socially assistive robots
  case study, Wirel. Pers. Commun. 117~(1) (2021) 261--280.

\bibitem{Cybenko}
G.~Cybenko, Approximations by superpositions of sigmoidal functions,
  Mathematics of Control, Signals, and Systems 2~(4) (1989) 303–314.
\newblock \href {https://doi.org/10.1007/BF02551274}
  {\path{doi:10.1007/BF02551274}}.

\bibitem{bib:GelenbeApprox1999}
E.~Gelenbe, Z.-H. Mao, Y.~Li, Function approximation with spiked random
  networks, IEEE Transactions on Neural Networks 10~(1) (1999) 3--9.

\bibitem{HarrisonPitel}
P.~G. Harrison, E.~Pitel, Response time distributions in tandem g-networks,
  Journal of Applied Probability 32~(1) (1995) 224–--246.
\newblock \href {https://doi.org/10.2307/3214932} {\path{doi:10.2307/3214932}}.

\bibitem{Harrison03}
P.~G. Harrison, \href{http://doi.acm.org/10.1145/959143.959144}{G-networks with
  propagating resets via {RCAT}}, {SIGMETRICS} Performance Evaluation Review
  31~(2) (2003) 3--5.
\newline\urlprefix\url{http://doi.acm.org/10.1145/959143.959144}

\bibitem{Fourneau13}
J.-M. Fourneau, K.~Wolter, P.~Reinecke, T.~Krau{\ss}, A.~Danilkina,
  \href{http://doi.acm.org/10.1145/2479871.2479880}{Multiple class g-networks
  with restart}, in: {ACM/SPEC} International Conference on Performance
  Engineering, ICPE'13, Prague, Czech Republic - April 21 - 24, 2013, 2013, pp.
  39--50.
\newline\urlprefix\url{http://doi.acm.org/10.1145/2479871.2479880}

\bibitem{bib:TIMOTHEOU2009}
S.~Timotheou, A novel weight initialization method for the random neural
  network, Neurocomputing 73~(1) (2009) 160 -- 168.

\end{thebibliography}

\end{document}